\documentclass[sigplan, 10pt]{acmart}
\setcopyright{none}
\def\BibTeX{{\rm B\kern-.05em{\sc i\kern-.025em b}\kern-.08emT\kern-.1667em\lower.7ex\hbox{E}\kern-.125emX}}
    
\usepackage{nicefrac}
\usepackage{siunitx}
\usepackage{array,framed}
\usepackage{booktabs}
\usepackage{
  color,
  float,
  epsfig,
  wrapfig,
  graphics,
  graphicx,
  subcaption
}
\usepackage{textcomp}
\usepackage{setspace}
\usepackage{latexsym,fancyhdr,url}
\usepackage{enumerate}
\usepackage{algorithm}
\usepackage{algorithmic}
\usepackage{graphics}
\usepackage{xparse} 
\usepackage{xspace}
\usepackage{multirow}
\usepackage{csvsimple}
\usepackage{balance}
\usepackage{microtype}

\makeatletter
\renewcommand{\ALG@name}{Algorithm}  
\renewcommand{\fnum@algorithm}{\ALG@name~\thealgorithm.}  
\makeatother

\makeatletter

\newcommand{\Rmnum}[1]{\expandafter\@slowromancap\romannumeral #1@}
\makeatother

\usepackage{
  tikz,
  pgfplots,
  pgfplotstable
}
\usepackage{hyperref}
\usepackage{xurl}

\usetikzlibrary{
  shapes.geometric,
  arrows,
  external,
  pgfplots.groupplots,
  matrix
}

\pgfplotsset{compat=1.9}

\usepackage{mathtools,}

\DeclareMathAlphabet{\mathcal}{OMS}{cmsy}{m}{n}

\DeclareGraphicsExtensions{%
    .png,.PNG,%
    .pdf,.PDF,%
    .jpg,.mps,.jpeg,.jbig2,.jb2,.JPG,.JPEG,.JBIG2,.JB2}

\usepackage{xparse}
\newcommand{\bnm}{\begin{newmath}}
\newcommand{\enm}{\end{newmath}}

\newcommand{\bea}{\begin{eqnarray*}}%
\newcommand{\eea}{\end{eqnarray*}}%

\newcommand{\bne}{\begin{newequation}}
\newcommand{\ene}{\end{newequation}}

\newcommand{\bal}{\begin{newalign}}
\newcommand{\eal}{\end{newalign}}

\newenvironment{newalign}{\begin{align}%
\setlength{\abovedisplayskip}{4pt}%
\setlength{\belowdisplayskip}{4pt}%
\setlength{\abovedisplayshortskip}{6pt}%
\setlength{\belowdisplayshortskip}{6pt} }{\end{align}}

\newenvironment{newmath}{\begin{displaymath}%
\setlength{\abovedisplayskip}{4pt}%
\setlength{\belowdisplayskip}{4pt}%
\setlength{\abovedisplayshortskip}{6pt}%
\setlength{\belowdisplayshortskip}{6pt} }{\end{displaymath}}

\newenvironment{newequation}{\begin{equation}%
\setlength{\abovedisplayskip}{4pt}%
\setlength{\belowdisplayskip}{4pt}%
\setlength{\abovedisplayshortskip}{6pt}%
\setlength{\belowdisplayshortskip}{6pt} }{\end{equation}}

\newcounter{ctr}

\newcounter{mytable}
\def\mytable{\begin{centering}\refstepcounter{mytable}}
\def\endmytable{\end{centering}}

\newcounter{myfig}
\def\myfig{\begin{centering}\refstepcounter{myfig}}
\def\endmyfig{\end{centering}}

\newlength{\saveparindent}
\newlength{\saveparskip}
\newcommand{\E}{{\rm I\kern-.3em E}}

\renewcommand{\eqref}[1]{\mbox{Equation~(\ref{#1})}}

\def \part {part}

\renewcommand{\paragraph}[1]{\vspace*{6pt}\noindent\textbf{#1}\;}

\def \blackslug{\hbox{\hskip 1pt \vrule width 4pt height 8pt
    depth 1.5pt \hskip 1pt}}
\def \qed{\quad\blackslug\lower 8.5pt\null\par}

\newcounter{mynote}[section]

\newcommand\ignore[1]{}

\newcounter{rcnote}[section]

\newcounter{mrnote}[section]

\newcounter{fknote}[section]

\newcounter{anote}[section]

\DeclareMathSymbol{\mlq}{\mathord}{operators}{``}
\DeclareMathSymbol{\mrq}{\mathord}{operators}{`'}

\newcommand{\rhf}[2]{R_{f, \gamma}}

\DeclareDocumentCommand{\edist}{o o}{
  \ensuremath{
    \IfNoValueTF{#1}{{d}}{{\sf d}(#1,#2)}
  }
}

\newcommand{\olrk}[1]{\ifx\nursymbol#1\else\!\!\mskip4.5mu plus 0.5mu\left(\mskip0.5mu plus0.5mu #1\mskip1.5mu plus0.5mu \right)\fi}

\NewDocumentCommand{\indseq}{ O{1} O{r} }{{#1}\ldots {#2}}

\begin{document}
\fancyhead{}
\def\thetitle{Deduplication-while-Training: A Resilient Paradigm for Privacy-Preserving Cross-Client Deduplication in Federated Learning}
\title{\thetitle}

\author{Rongxi Wang}
\affiliation{
	\institution{College of Cryptology and Cyber Science, Nankai University}
	\city{Tianjin}
	\country{China}
}

\author{Guanxiong Ha}
\affiliation{
	\institution{College of Cryptology and Cyber Science, Nankai University}
	\city{Tianjin}
	\country{China}
}

\author{Chunfu Jia}
\affiliation{
	\institution{College of Cryptology and Cyber Science, Nankai University}
	\city{Tianjin}
	\country{China}
}

\author{Yongsheng Lin}
\affiliation{
	\institution{College of Cryptology and Cyber Science, Nankai University}
	\city{Tianjin}
	\country{China}
}

\author{Minfen Gao}
\affiliation{
	\institution{School of Mathematical Science, Nankai University}
	\city{Tianjin}
	\country{China}
}

\author{Hanmiaomiao Wang}
\affiliation{
	\institution{College of Cryptology and Cyber Science, Nankai University}
	\city{Tianjin}
	\country{China}
}

\date{}

\setlength{\abovedisplayskip}{6pt}   
\setlength{\belowdisplayskip}{6pt}

\begin{abstract}
	
\begin{sloppypar} 
Cross-client duplicate data in large language model training corpora degrades the efficiency of federated learning (FL) while exacerbating model memorization and privacy risks. 
Privacy-preserving cross-client deduplication effectively mitigates this issue by eliminating duplicate training data. 
However, existing schemes all follow a "Deduplication-before-Training" paradigm. 
This serially coupled paradigm incurs high fault-tolerance costs and lacks support for dynamic client joining. 

To this end, we propose an unexplored paradigm called "Deduplication-while-Training (DwT)", which enables concurrent deduplication and training. 	
DwT transforms cross-client deduplication from a one-time, globally synchronous preprocessing operation into a continuous online service with state management, concurrent claiming, and failure recovery. 	
By enabling state synchronization and task takeover, it minimizes the impact of client disconnections on the overall training progress while supporting the dynamic joining of clients. 
We design DwT-FL, a privacy-preserving deduplication system, to support DwT. 
By designing a concurrent state-claim mechanism and a hot-cold dual-queue scheduling strategy, DwT-FL enables the parallel execution of secure deduplication and model training, while effectively handling client disconnections and dynamic joins. 
Experimental evaluations demonstrate that, compared to the state-of-the-art scheme, DwT-FL significantly reduces the time overhead of failure recovery and dynamic joining by up to $93.04\%$ and $94.18\%$, respectively. 
This provides an efficient and elastic concurrent deduplication scheme for dynamic and unstable FL environments.

\end{sloppypar}

\end{abstract}

\maketitle

\section{Introduction}
\label{sec:intro}

The performance of large language models (LLMs) \cite{Ouyang0JAWMZASR22, abs-2307-09288, 0001LV23, abs-2303-08774, abs-2412-19437} depends strongly on the scale and quality of their training data. 
However, large real-world text corpora commonly contain noise, bias, and substantial duplicate content \cite{AbadiDS25}. 
Duplicated sequences can cause a model to overemphasize a limited set of sample patterns, weaken generalization, and increase the risk that training data are memorized and reproduced. 
Prior work shows that duplicate training data are closely associated with model memorization and can further increase the success of privacy attacks \cite{CarliniIJLTZ23}, including membership inference attacks \cite{abs-2206-03317, MatternMJSSB23} and data extraction attacks \cite{NasrRCHJCICTL25, CarliniTWJHLRBS21, abs-2310-16152, FowlGRWCGG23, BalunovicD0V22}. 
Duplicate samples also incur unnecessary computation and storage overhead, lengthen training time, and increase energy costs. 
Therefore, \textit{the deduplication of training data} serves as a crucial step for enhancing model quality, privacy, and resource utilization in LLM training \cite{LeeINZECC22, BenderGMS21, StrubellGM19, patterson2021carbon}.

\begin{sloppypar} 
Federated learning (FL) \cite{0001L0025, ChenXZM25, GanLALLZ025, McMahanMRHA17} provides a practical approach for collaborative training on private data distributed across users or devices. 
Clients train models locally and upload only model updates to a central server, avoiding direct exchange of raw data \cite{McMahanMRA16}. 
FL has been applied to mobile input prediction \cite{abs-1811-03604}, healthcare \cite{AntunesCKYE22}, smart cities \cite{JiangKOS20}, and edge computing \cite{XiaYTWL21}. 
However, different clients in FL can hold identical text fragments, samples, or data records, and local-only deduplication cannot identify duplicates across clients \cite{abs-1811-03604, AbadiDS25, YeLSZDCYDY25}. 
To perform cross-client deduplication while preserving data privacy, the system must detect duplicates without exposing the original data content. 
Thus, \textit{privacy-preserving cross-client deduplication represents a critical challenge in FL}.

Existing privacy-preserving cross-client deduplication schemes \cite{AbadiDS25, YeLSZDCYDY25} rely mainly on trusted execution environments (TEEs) or cryptographic techniques to identify duplicates without directly disclosing data content. 
These approaches organize deduplication and training through a \textit{Deduplication-before-Training (DbT)} paradigm. 
DbT requires all participants to complete a duplicate-detection protocol synchronously before clients train on the deduplicated data. 
By treating deduplication as a preprocessing step, DbT ensures that clients avoid training on duplicate data, thereby reducing training costs and mitigating model memorization. 
However, this serially coupled design paradigm has two limitations. 
1) \textit{Significant fault-tolerance costs}: If some clients unexpectedly disconnect during deduplication or training, the system commonly needs to roll back its state and re-execute the prescribed protocol, introducing significant  additional communication and computation overhead.
2) \textit{No support for dynamic client joining}: DbT requires all clients to uniformly complete deduplication before jointly entering the training phase. 
When a new client joins, it is required to execute the deduplication protocol with all existing clients, which introduces high communication and computation overhead and restricts the system's scalability.
\end{sloppypar}

The limitations of DbT motivate us to explore a novel paradigm called \textit{Deduplication-while-Training (DwT)} paradigm, which enables concurrent deduplication and training. 
Rather than requiring all participants to establish a common synchronized state before duplicate detection and training, DwT reduces the granularity of these operations to individual data items. 
Clients can come online asynchronously. 
After completing duplicate detection through interaction with the FL server, a client can immediately train on data for which it has obtained training rights, without waiting for other participants to be online or for global deduplication to finish. 
The challenge of implementing DwT lies in ensuring that duplicate content in the training data across all clients is trained exactly once, while providing both privacy preservation and high system performance within complex network environments characterized by unexpected client dropouts and dynamic joining. 

\begin{sloppypar} 
We design DwT-FL, a system that implements the DwT paradigm. 
By designing \textit{a concurrent state-claim mechanism} based on compare-and-swap (CAS) and \textit{a hot-cold dual-queue scheduling strategy}, DwT-FL enables privacy-preserving cross-client duplicate detection and client-local training to proceed in parallel. 
It consists of clients, an aggregation server ($\mathcal{AS}$), and a key server ($\mathcal{KS}$). 
Each client first runs an Oblivious Pseudorandom Function (OPRF) protocol \cite{BurnsMRSV17} with $\mathcal{KS}$ to generate protected duplicate detection data tags for local training data. 
$\mathcal{AS}$ performs duplicate detection and model-parameter aggregation.
Additionally, it maintains a critical data structure for implementing DwT, the \textit{state table}, which records each data item's tag, its current state, and owner information. 
The data states are $\mathtt{EMPTY}$, $\mathtt{PENDING}$, and $\mathtt{COMMITTED}$, indicating that the data item is unclaimed, being trained, and successfully trained in the current round, respectively. 

When multiple clients submit the same tag, $\mathcal{AS}$ identifies cross-client duplicate data. 
$\mathcal{AS}$ uses lock-free updates and CAS operations to handle concurrent requests and ensure a unique training right for each data item. 
A client that successfully claims a training right receives a training identifier, moves the data item to its local hot queue, and trains it. 
Other clients that own the data item receive a deduplication identifier and retain the data item in a cold queue rather than deleting it. 
If a training client $U$ disconnects unexpectedly, $\mathcal{AS}$ uses the heartbeat mechanism and the owner information recorded in the state table to schedule training rights. 
$\mathcal{AS}$ transfers the training rights of the data to other clients that possess the same data as client $U$. 
These clients then move the data from their cold queues to their hot queues and execute the training, preventing data loss caused by client disconnections. 
When a new client joins during system execution, $\mathcal{AS}$ incrementally updates the state table without interrupting existing clients or rerunning a global deduplication protocol. 
In addition, $\mathcal{AS}$ retains historical trainers and disconnection information during cross-round scheduling to support efficient and stable multi-round federated training.
\end{sloppypar}

The central advantage of DwT is that it transforms cross-client deduplication in FL from a one-time, globally synchronized preprocessing operation into \textit{a continuously available service with state management, concurrent claiming, and failure recovery}. 
This paradigm allows clients that have completed duplicate detection to enter training promptly, reducing idle waiting. 
It reduces the impact of client disconnections on overall training progress through state recovery and task takeover, and incrementally handles dynamically joined clients without repeating the complete deduplication process. 
Without changing the privacy objective that cross-client data content remains hidden, DwT coordinates the execution of data deduplication and federated training and provides a more resilient system organization for FL environments with unstable networks and dynamic participants.

In summary, our contributions are as follows:

\begin{itemize}
	\item We propose the DwT paradigm, which for the first time transforms cross-client duplicate detection in FL from a one-time, globally synchronized preprocessing operation into a continuously available service with state management, concurrent claiming, and failure recovery. 
	The paradigm handles client disconnections and supports dynamic client joining.
	
	\item We design DwT-FL to support DwT. 
	Through a CAS-based concurrent state-claim mechanism and a hot-cold dual-queue scheduling strategy, DwT-FL enables privacy-preserving cross-client data deduplication and client-local training to proceed in parallel, while ensuring high scalability and resilience to client disconnections.
	
	\item We conduct a systematic evaluation of DwT-FL along the dimensions of system performance, and fault tolerance. 
	Experimental results show that, compared to the state-of-the-art scheme, DwT-FL reduces the failure recovery time by up to $93.04\%$, and decreases the time overhead for dynamic client joining by up to $94.18\%$ and $78.87\%$ in the deduplication and training phases, respectively. 
	Furthermore, we make the source code of DwT-FL publicly available\footnote{\url{https://anonymous.4open.science/r/DwT-FL}}.
\end{itemize}

\section{Background and Motivation}
\label{sec:background}

\begin{figure*}[t]
	\centering
	\includegraphics[width=0.8\textwidth]{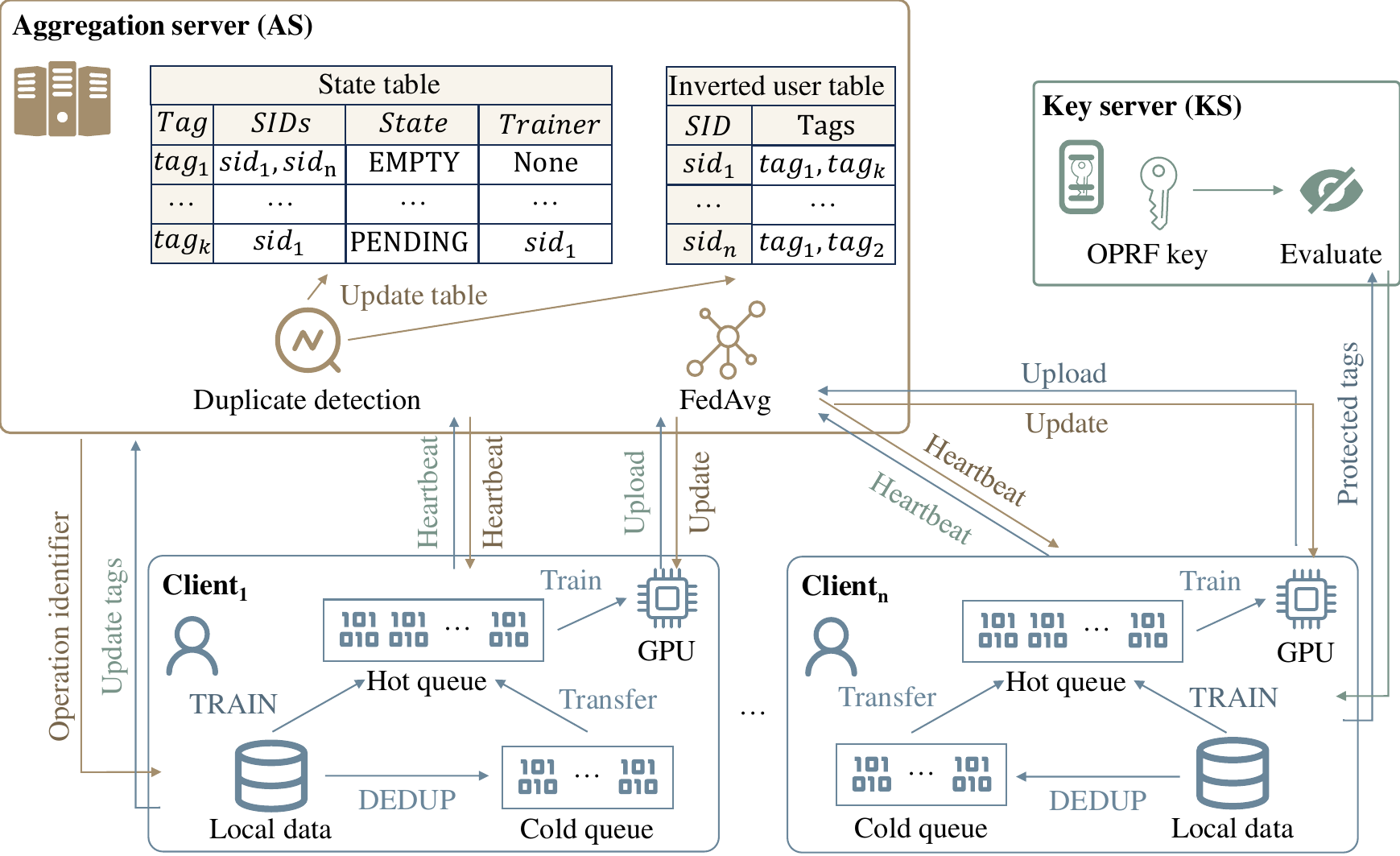}
	\caption{Architecture of DwT-FL.}
	\label{fig:Architecture}
\end{figure*}

\subsection{Federated Learning}

FL is a distributed machine learning paradigm \cite{0001L0025, ChenXZM25, GanLALLZ025, McMahanMRHA17}. 
In conventional model training, data are collected at a central server; in FL, data remain on clients' personal devices \cite{McMahanMRA16}. 
A typical FL system trains a model through multiple rounds of interaction between a server and multiple clients. 
In each round, the central server distributes the current global model parameters to participating clients. 
Each client uses locally stored private data to execute an optimization algorithm, such as stochastic gradient descent, and computes a local model update, such as a gradient. 
Clients upload only model updates to the server, which aggregates the collected updates with an algorithm such as federated averaging (FedAvg \cite{McMahanMRA16}), produces a new global model, and begins the next round. 
Under this architecture, raw data remain on local devices, avoiding direct sharing or centralized storage of private data. 
Notably, our scheme is independent of the specific FL approach employed.

By retaining raw data locally during training, FL mitigates privacy risks associated with conventional centralized training. 
However, because clients do not upload their data to the server, duplicate training records are difficult to detect. 
Training the same data repeatedly at multiple clients incurs additional computation and communication overhead. 
More importantly, repeated training can cause overfitting to duplicate data and make the model more susceptible to security threats such as membership inference attacks \cite{abs-2206-03317, MatternMJSSB23} and data extraction attacks \cite{NasrRCHJCICTL25, CarliniTWJHLRBS21, abs-2310-16152, FowlGRWCGG23, BalunovicD0V22}. Privacy-preserving cross-client data deduplication is therefore important for FL.

\subsection{Existing Privacy-Preserving Cross-Client Deduplication Schemes for FL}

Existing schemes transform global deduplication among FL participants into multi-party cryptographic interaction protocols, thereby detecting duplicate data without directly exposing plaintext. 
Abadi et al. \cite{AbadiDS25} propose EP-MPD, a privacy-preserving multi-party deduplication protocol based on a binary-tree structure. 
With assistance from TEEs, clients are paired at each tree level and use group private set intersection (PSI) \cite{FreedmanNP04, abs-2301-03889, AbadiDMT22} based on symmetric encryption or OPRF to identify duplicate data, after which one endpoint removes the duplicate. 
This interaction recurses upward through the tree until the root is reached, eliminating cross-client data redundancy. 
Ye et al. \cite{YeLSZDCYDY25} further consider the frequency of duplicate data, adjust training weights according to that frequency, and decompose cross-client duplicate identification into multiple rounds of PSI-based secure two-party computation. 

\begin{sloppypar} 
However, all existing privacy-preserving cross-client deduplication schemes \cite{AbadiDS25, YeLSZDCYDY25} for FL follow the DbT paradigm. 
As described in Section \ref{sec:intro}, this paradigm treats deduplication as preprocessing before training and is vulnerable to unexpected client disconnections. 
If a client disconnects during deduplication or training, the system must rerun the complete protocol, introducing substantial communication and computation overhead. 
Moreover, DbT does not support dynamic client joining and therefore has limited scalability.
\end{sloppypar}
\section{System Overview}
\label{sec:overview}

\subsection{Architecture}

\begin{sloppypar} 
As shown in Fig. \ref{fig:Architecture}, DwT-FL consists of a $\mathcal{AS}$, a $\mathcal{KS}$, and multiple clients. 
$\mathcal{KS}$ generates and securely stores a system-wide OPRF key, and it assists clients in generating protected data tags used for duplicate detection. 
$\mathcal{AS}$ performs deduplication on the training data of clients, receives and aggregates federated model parameters, and manages client sessions and data item states during deduplication and training. 
Clients in the system interact with $\mathcal{KS}$ to generate a duplicate detection tag for each training data item, and they send these tags to $\mathcal{AS}$ to execute cross-client duplicate detection. 

After $\mathcal{AS}$ executes duplicate detection, it determines a unique trainer for each training data item based on a concurrent state-claim mechanism. 
The client that acquires the training right places the data into the hot queue and performs training immediately; other clients place the data into the cold queue for temporary storage, and this data is used for failure recovery. 
After clients complete the training, they send model updates to $\mathcal{AS}$, and after $\mathcal{AS}$ aggregates these updates, it obtains a new global model and prepares to start the next training round. 
If a client unexpectedly disconnects, other clients can submit their own requests to replace the disconnected client and continue training, and this does not impact the overall performance of the system. 
Furthermore, the system incrementally handles dynamically joined clients, and it does not affect other clients. 

\paragraph{Remark.}
It should be noted that the DwT paradigm does not restrict the adopted deduplication methods (for example, exact deduplication or MinHash-based fuzzy deduplication). 
The primary focus of DwT-FL is the efficient implementation of failure recovery and the support for dynamic client joining. 
Therefore, the deduplication method primarily implemented by DwT-FL is exact deduplication.
If fuzzy deduplication is required, the exact deduplication method in DwT-FL can simply be replaced with existing schemes \cite{abs-2303-09540, TirumalaSAM23, JiangYCCWCM23, SongHZXJ24}. 
\end{sloppypar}

\subsection{Threat Model}

\begin{sloppypar} 
We assume that $\mathcal{AS}$ is honest-but-curious. 
That is, it follows the prescribed protocol but attempts to infer user data from received information. 
$\mathcal{KS}$ is also honest-but-curious, meaning that it correctly evaluates OPRF but may attempt to infer information about data from client inputs. 
Following existing schemes \cite{KeelveedhiBR13, HaJHCLJ24, Lai0RCMS18, Lai0SC17}, we assume that $\mathcal{AS}$ and $\mathcal{KS}$ do not collude. 
This assumption is feasible in practical scenarios \cite{HaTC26}. 
Typically, $\mathcal{AS}$ and $\mathcal{KS}$ are not operated by the same service provider. 
Model aggregation can be hosted by a public cloud provider with sufficient training resources, while cryptographic services can be hosted by an independent third-party security organization. 
This deployment strategy not only leverages the powerful computational capacity of the public cloud but also strengthens data security, effectively circumventing the risk of collusion between $\mathcal{AS}$ and $\mathcal{KS}$. 
We assume that clients are honest and do not collude with $\mathcal{AS}$ or $\mathcal{KS}$, but may disconnect or delay their participation in FL because of network conditions. 
All communication channels among clients, $\mathcal{AS}$, and $\mathcal{KS}$ are authenticated, for example through TLS.
\end{sloppypar}

\paragraph{Remark.}
Similar to the scheme proposed by Abadi et al. \cite{AbadiDS25}, our design deploys a third-party $\mathcal{KS}$ to strengthen security. 
Although Ye et al. \cite{YeLSZDCYDY25} do not rely on a third-party server, they require clients to execute pairwise PSI while all clients remain online and can implement only the DbT paradigm. 
When a client disconnects, these schemes must roll back state and rerun the prescribed protocol, which introduces substantial communication and computation overhead. 
They also do not support dynamic client joining.

\section{Detailed Method}
\label{sec:method}

\subsection{Design Goals}

\begin{sloppypar} 
DwT-FL targets federated learning scenarios characterized by unstable network connections and dynamically participating clients. 
Its objective is to achieve the parallel execution of cross-client duplicate data identification and client-local training while preserving data privacy. 
Specifically, DwT-FL has three design goals. 
1) \textit{Data privacy preservation}. 
During cross-client duplicate detection, the system needs to ensure that the local plaintext data of all participants is not disclosed.
2) \textit{Resilience to client dropouts}. 
When some clients experience network disconnections during deduplication or training, the system needs to avoid blocking the deduplication or training of other online clients. 
3) \textit{Support for dynamic client joining}. 
To accommodate dynamically participating clients, the system needs to process joining requests asynchronously. 
Specifically, the duplicate detection and training of new clients should not interrupt the ongoing execution of existing online clients, nor should it require a global re-execution of the deduplication protocol. 
\end{sloppypar}

\subsection{Construction of DwT-FL}

\begin{sloppypar} 
\paragraph{System initialization.} 
During system initialization, $\mathcal{KS}$ generates a security parameter $\lambda$ and, based on this parameter, generates and securely stores an OPRF key $k \in Z_p^*$, where $Z_p^*$ denotes the multiplicative group of integers modulo a large prime $p$. 
To support efficient duplicate detection and failure recovery under concurrent requests from multiple clients, $\mathcal{AS}$ initializes \textit{a lock-free bidirectional concurrent hash index}. 
The index contains a state table and an inverted user table. 
The state table is keyed by $fp_M$, where $fp_M$ denotes the protected tag of data item $M$. 
Each entry records the current processing state of the corresponding data item ($\mathtt{EMPTY}$, $\mathtt{PENDING}$, or $\mathtt{COMMITTED}$), the set of data owners (a list of $SID$s, where each $SID$ is the identifier of a client owning the data), and the client selected to train the data item in the current round. 
$\mathtt{EMPTY}$ indicates that the data item has not been claimed, $\mathtt{PENDING}$ indicates that it is being trained, and $\mathtt{COMMITTED}$ indicates that its training has completed successfully in the current round. 
The inverted user table is keyed by $SID$ and records the set of protected tags for all data owned by that client.
\end{sloppypar}

When a new client (e.g., $U_i$) joins the network, $\mathcal{AS}$ assigns it a globally unique $SID_i$ and establishes \textit{a heartbeat mechanism}. 
Specifically, the system configures a fixed heartbeat interval $\Delta t$ and a timeout threshold $\tau$. 
After successfully joining the network and obtaining $SID_i$, the client starts a background timer that continuously sends heartbeat messages to $\mathcal{AS}$ every $\Delta t$. 
$\mathcal{AS}$ maintains an independent timer for each online client and resets the corresponding timer upon receiving a valid heartbeat. 
If a timer has not been refreshed for longer than $\tau$, $\mathcal{AS}$ marks that client as disconnected. 
During subsequent deduplication and training, $\mathcal{AS}$ determines whether a client has disconnected based on whether this timer exceeds $\tau$ and triggers the failure recovery mechanism when this condition is met (detailed in the \textit{failure recovery} paragraph).

\paragraph{Protected-tag generation.} 
Suppose that client $U_i$ joins the network with a local plaintext dataset $D_i$. 
The dataset contains both records duplicated at other clients and records unique to $U_i$. 
$U_i$ interacts with $\mathcal{KS}$ to generate a protected tag for each data item in $D_i$. 
For a training data item $m_j \in D_i$, $U_i$ first computes its cryptographic hash $H(m_j)$, samples a random value $r \leftarrow \{0, 1\}^\lambda$, and sends the blinded value $H(m_j)^r$ to $\mathcal{KS}$. 
$\mathcal{KS}$ evaluates $(H(m_j)^r)^k$ using the OPRF key $k$ and returns the result. 
$U_i$ then computes $((H(m_j)^r)^k)^{r^{-1}} = H(m_j)^k$ locally and obtains the protected tag $fp_{m_j} = H(m_j)^k$. 
After applying this procedure to every data item in $D_i$, $U_i$ obtains the tag set $FP_i=\{fp_{m_1},fp_{m_2},\cdots,fp_{m_n}\}$, where $n$ is the number of local data items. 
Subsequently, $U_i$ sends $FP_i$ together with its identifier $SID_i$ to the $\mathcal{AS}$ to initiate a deduplication request.

\begin{sloppypar} 
\paragraph{Bidirectional-index update.}
After receiving a client deduplication request, $\mathcal{AS}$ performs a lock-free update of the bidirectional index. 
For a newly joined client $U_i$, $\mathcal{AS}$ initializes an empty tag set for $U_i$ in the inverted user table and appends $FP_i$. 
Subsequently, $\mathcal{AS}$ updates the state table based on $FP_i$. 
$\mathcal{AS}$ checks whether the tag in $FP_i$ has already been stored in the state table. 
If it has been stored, it indicates that this data item is duplicated; otherwise, it is unique.
$\mathcal{AS}$ iterates through each $fp_{m_j}$ in $FP_i$ and performs different operations depending on whether the data is a duplicate. 
Specifically, for the tags corresponding to data unique to $U_i$, $\mathcal{AS}$ creates new entries in the state table, initializes their states to $\mathtt{EMPTY}$, and records $SID_i$ as the sole owner. 
For duplicate data, two cases arise depending on whether other clients in the network concurrently submit the same tag. 
\begin{enumerate}
	\item \textit{Concurrent submission.} 
	If another client $U_j$ concurrently submits the corresponding tag $fp_{m_j}$ to $\mathcal{AS}$, to prevent concurrency conflicts, $\mathcal{AS}$ performs a lock-free insertion of the tag. 
	This operation ensures that under concurrent contention, only one thread successfully allocates memory for the tag and initializes its state to $\mathtt{EMPTY}$. 
	The remaining threads directly obtain this memory reference to subsequently compete for the data's training right. 
	Following this, the identifiers of both $U_i$ and $U_j$ are safely appended to the data's owner set.
	
	\item \textit{Non-concurrent submission.}
	If no other client concurrently submits the tag for this data to $\mathcal{AS}$, the system directly and safely appends $U_i$'s identifier to the data's owner set.
\end{enumerate}

It is worth noting that when $\mathcal{AS}$ successfully allocates memory space for the tag, it does not directly assign the training right to the corresponding client. 
The system decouples data entry creation from training-right allocation to handle client disconnections during the deduplication phase. 
If $\mathcal{AS}$ directly binds the training right when allocating memory, the data would be invalidly occupied if the client disconnects during this stage. 
By deferring the allocation of training rights until the subsequent competition using CAS operations, the system ensures that even if the client that successfully allocated the memory space disconnects during deduplication, the data's state remains $\mathtt{EMPTY}$, thereby allowing other clients to normally claim the training right. 

\begin{figure}[t]
	\centering         
	\includegraphics[width=1.0\linewidth]{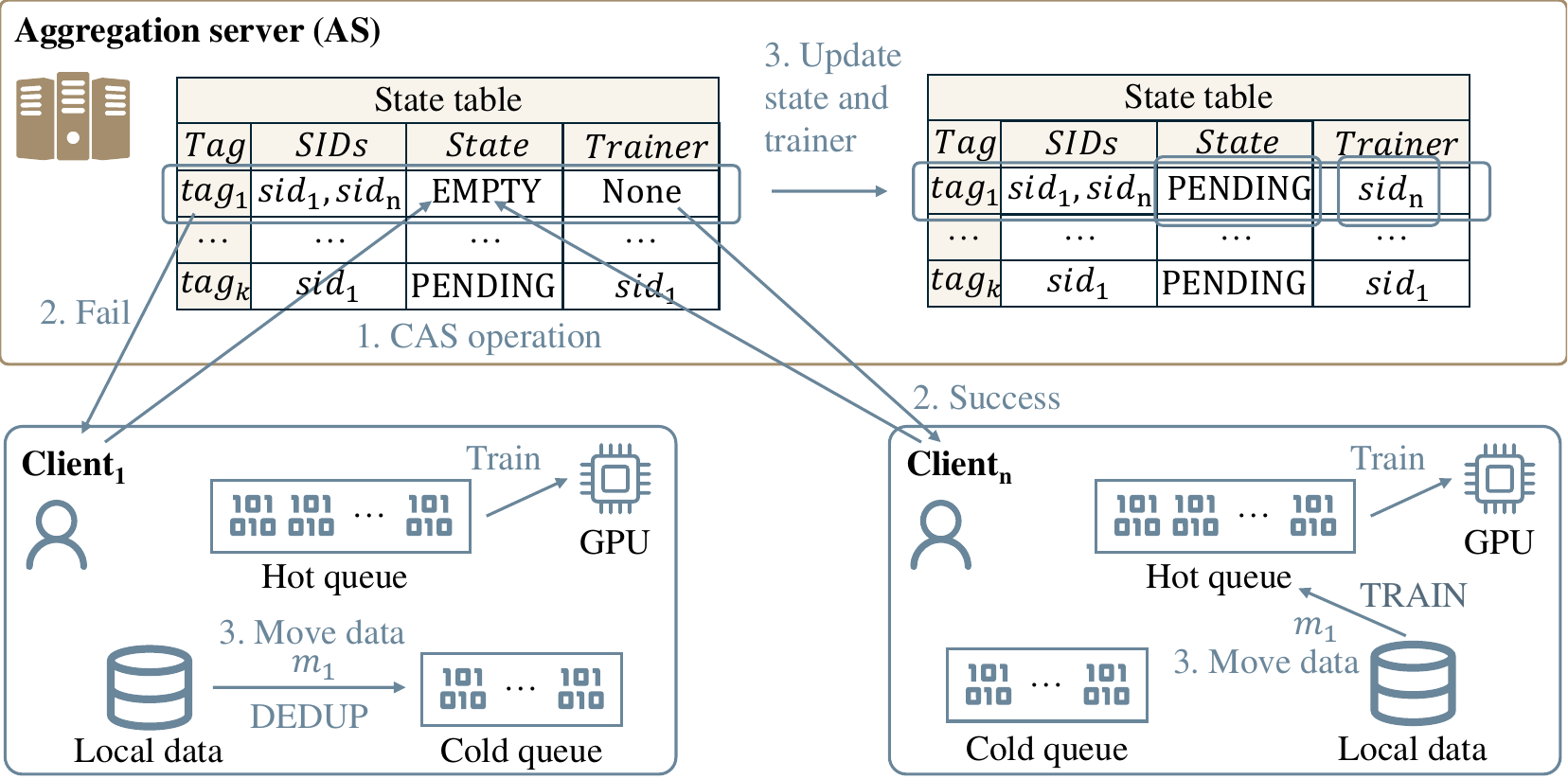} 
	\caption{Unique trainer selection for contested data.}
	\label{fig:Unique trainer selection}   
\end{figure}

\paragraph{Unique trainer selection.}
After the bidirectional index is updated, $\mathcal{AS}$ determines the unique trainer for every data item submitted by $U_i$ in the current round through \textit{the concurrent state-claim mechanism}. 
Under optimistic concurrency control \cite{KungR81}, when attempting to acquire the training right, the thread processing the client request first performs a lock-free initial memory read of the data entry state. 
If, during this initial memory read phase, the processing thread discovers that the data entry state is already $\mathtt{PENDING}$ or $\mathtt{COMMITTED}$ (indicating that the data has been successfully claimed by another client that arrived earlier), $\mathcal{AS}$ determines that the client has failed to claim this tag. 

Conversely, if the data entry state is $\mathtt{EMPTY}$, two cases arise. 
1) For contested data, where multiple concurrent threads (e.g., $U_i$ and $U_j$) read the entry state as $\mathtt{EMPTY}$ simultaneously, $\mathcal{AS}$ requires these threads to execute \textit{a CAS operation} on the data entry state variable, as shown in Fig. \ref{fig:Unique trainer selection}. 
During execution, the CAS operation receives three parameters: the memory address of the data entry state, the expected old value ($\mathtt{EMPTY}$), and the new value to be written ($\mathtt{PENDING}$). 
Using these parameters, CAS atomically compares the actual value at the memory address with the expected old value. 
If they match, it proves that the data entry has not been modified by any other concurrent threads. 
CAS then writes the new value $\mathtt{PENDING}$ into memory and binds the current-round trainer field to the thread's corresponding $SID_i$ (assuming $U_i$'s thread executes the fastest), resulting in a successful claim. 
If they do not match (e.g., the actual value has already been modified to $\mathtt{PENDING}$ by $U_i$'s thread), the CAS operation directly aborts the modification. 
2) For uncontested data (including unique data and duplicate data without concurrent access), $U_i$'s processing thread executes the aforementioned CAS operation and can successfully complete the state binding via the CAS operation. 
\end{sloppypar}

\paragraph{Training and deduplication.}
The system implements \textit{the hot-cold dual-queue scheduling strategy} to assign training rights and manage local data based on the duplicate detection results. 
After $\mathcal{AS}$ has processed all tags in a client deduplication request, it aggregates the results and returns a response message to that client.
This message contains a state mapping table that records each of the client's tags and its corresponding operation identifier. 
For successfully claimed tags (including unique data and successfully claimed duplicate data), the \texttt{TRAIN} operation identifier is attached. 
For unsuccessfully claimed tags, the \texttt{DEDUP} operation identifier is attached. 
Upon receiving this response message, the client can locate the plaintext data in its local dataset based on the state mapping table within the message, and route the data to a hot queue or a cold queue respectively according to the operation identifiers. 
Specifically, $U_i$ moves the plaintext data corresponding to \texttt{TRAIN} into its local hot queue and immediately begins training, without waiting for other clients in the network. 
Conversely, $U_j$ transfers the plaintext data corresponding to \texttt{DEDUP} into its local cold queue to suspend it, preparing for potential subsequent takeover. 
It is worth noting that $U_j$ does not delete the data that needs to be deduplicated; instead, it retains the data on its local disk to provide data support for the failure recovery mechanism that may occur later. 

\paragraph{Completion of a training round.}
After all clients complete local training in a round and upload model updates to $\mathcal{AS}$, $\mathcal{AS}$ aggregates the updates to obtain a new global model and prepares the next iteration. 
At the cross-round transition, $\mathcal{AS}$ traverses the bidirectional index and resets the processing state of every tag to $\mathtt{EMPTY}$ while retaining its owner set and the previous round's unique trainer record. 
This reset is necessary for correct state progression. 
Retaining $\mathtt{COMMITTED}$ would prevent $\mathcal{AS}$ from distinguishing whether a data item has been processed in the new round. 
Directly assigning $\mathtt{PENDING}$ across rounds would incorrectly treat a disconnected default trainer as active and delay recovery until heartbeat timeout. 
Resetting to $\mathtt{EMPTY}$ denotes that a data item is ready but has not yet begun training in the new round. 
The state changes to $\mathtt{PENDING}$ only after the subsequent trainer sends a heartbeat and $\mathcal{AS}$ encapsulates the \texttt{TRAIN} instruction in the heartbeat response. 
This mechanism ensures that state transitions are explicitly bound to the client's online interaction and enables $\mathcal{AS}$ to accurately track the current training progress. 

In a stable setting without client disconnections, to avoid repeated duplicate detection and concurrent contention, the training right for each data item in the new round is retained by default for the previous round's trainer. 
Specifically, at the beginning of the next iteration, $\mathcal{AS}$ directly issues operation identifiers in the heartbeat responses sent to each client based on the retained trainer records. 
For any given data item, $\mathcal{AS}$ directly encapsulates the \texttt{TRAIN} operation identifier in the heartbeat response of its previous round's trainer (e.g., $U_i$). 
If the data is a duplicate shared by multiple clients, $\mathcal{AS}$ encapsulates the \texttt{DEDUP} operation identifier in the heartbeat responses of the other clients holding this data (e.g., $U_j$). 
Through this cross-round state reuse mechanism, after receiving the new round's global model, clients can directly train or suspend data in their local queues based on the heartbeat responses. 
This avoids the overhead of repeating CAS contention in every round and ensures the efficiency of multi-round federated training. 

\begin{figure}[t]
	\centering         
	\includegraphics[width=0.9\linewidth]{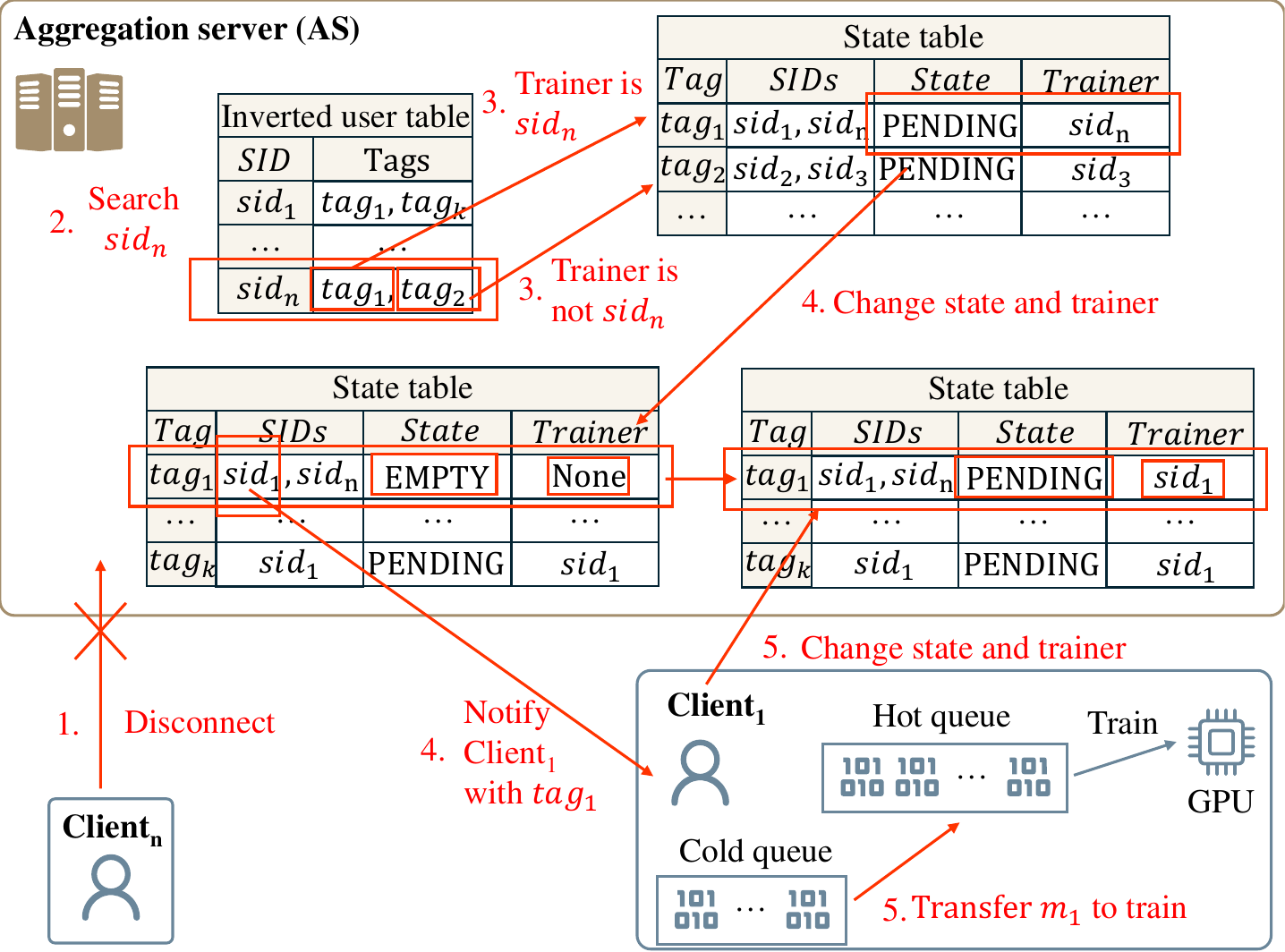} 
	\caption{Failure recovery in the training phase.}
	\label{fig:failure_recovery}   
	\vspace{-0.3cm}
\end{figure}

\paragraph{Failure recovery.}  
The fault-tolerance mechanism covers client disconnections during both the deduplication and training phases. 

1) \textit{Disconnection in the deduplication phase}. 
A client (e.g., $U_i$) unexpectedly disconnects during the deduplication phase (for example, while interacting with $\mathcal{KS}$ to execute OPRF, or before completely submitting its deduplication request to $\mathcal{AS}$). 
In the existing DbT paradigm, such local client disconnections typically cause protocol interruption, requiring the system to re-execute the complete deduplication process. 
In contrast, under the DwT paradigm, the system can ensure that the processing of this duplicate data remains unaffected without re-executing the protocol. 
Other clients holding the same data as $U_i$ (e.g., $U_j$) can still independently submit their tags to $\mathcal{AS}$. 
Because the data entry state remains $\mathtt{EMPTY}$, $U_j$ can successfully claim the training right through CAS and execute local training, thereby avoiding the problem of re-executing the deduplication procedure found in existing deduplication schemes \cite{AbadiDS25, YeLSZDCYDY25}.

\begin{sloppypar} 
2) \textit{Disconnection in the training phase}. 
As shown in Fig. \ref{fig:failure_recovery}, $U_i$ has successfully claimed the training right but disconnects due to network instability during training (or after $\mathcal{AS}$ has updated the state but before $U_i$ receives the response message).  
According to system settings, all clients must send heartbeats at a periodic interval of $\Delta t$. 
When the timer on $\mathcal{AS}$ detects that $U_i$ has not responded beyond the timeout threshold $\tau$, $\mathcal{AS}$ determines that the client has disconnected. 
At this time, the processing states of all $fp_m$ (including duplicate and unique data) assigned to $U_i$ for training stall in the $\mathtt{PENDING}$ state. 
$\mathcal{AS}$ first retrieves the tags owned by the disconnected client using the inverted user table, and resets a tag's stata from $\mathtt{PENDING}$ to $\mathtt{EMPTY}$ only if the disconnected client is the current trainer of that tag. 

Subsequently, $\mathcal{AS}$ queries the state table. 
For $U_i$'s unique data, since there are no other owners, its state remains $\mathtt{EMPTY}$ awaiting $U_i$'s reconnection. 
For duplicate data, $\mathcal{AS}$ retrieves other online owners (e.g., $U_j$) or new clients (e.g., $U_k$) that dynamically join the network and submit the same data at this exact time. 
When $U_j$ sends its next heartbeat, or when the new client $U_k$ initiates a deduplication request, $\mathcal{AS}$ encapsulates the $fp_m$ awaiting supplementary training in the response message and sends it back. 
Upon receiving the response, $U_j$ can transfer this plaintext data from its cold queue to its hot queue for supplementary training (if the newly joined $U_k$ successfully claims it, it directly adds the data to its hot queue for training). 
After completing the computation and uploading the model, $\mathcal{AS}$ updates the state of the corresponding $fp_m$ to $\mathtt{COMMITTED}$. 
Furthermore, $\mathcal{AS}$ caches the records of duplicate data shared between $U_i$ and other users to facilitate $U_i$ rejoining deduplication and training after restoring its network connection. 

\textit{Client reconnection.} 
If the disconnected client $U_i$ subsequently restores its network connection and rejoins the system, the system executes corresponding state synchronization process based on the phase in which $U_i$ disconnected. 
\begin{enumerate}
	\item \textit{Disconnection during deduplication.} 
	If client $U_i$ disconnected in deduplication phase, its subsequent act of restoring its network connection and re-initiating a deduplication request will not affect the normal operation of the system. 
	At this point, $\mathcal{AS}$ treats its request as a routine client request. 
	For data already claimed by other online clients and in the $\mathtt{PENDING}$ or $\mathtt{COMMITTED}$ state, $U_i$ directly receives the \texttt{DEDUP} identifier. 
	For data still in the $\mathtt{EMPTY}$ state, $U_i$ continues to attempt claiming the training right via CAS operations. 
	
	\item \textit{Disconnection during training.} 
	If $U_i$ disconnected after acquiring the training right, when it reconnects and attempts to submit training parameters, $\mathcal{AS}$ rejects this submission. 
	$\mathcal{AS}$ encapsulates the previously recorded tags corresponding to the duplicate data shared between $U_i$ and other users into the response message, and attaches the \texttt{DEDUP} identifier to them. 
	This instructs $U_i$ to move this data that has already been taken over into the cold queue. 
	Simultaneously, the local training progress executed by $U_i$ prior to disconnection on this portion of duplicate data is discarded. 
	The system requires $U_i$ to re-initiate training only on the untaken data (unique data) for which it regains the training right. 
\end{enumerate}

\paragraph{Dynamic client joining.}
When a new client joins, it generates tags and sends a deduplication request to $\mathcal{AS}$. 
$\mathcal{AS}$ only needs to perform an append update to the lock-free bidirectional index. 
During the allocation of training rights, if a newly submitted tag has already been claimed by an existing online client (i.e., reading a state of $\mathtt{PENDING}$ or $\mathtt{COMMITTED}$), $\mathcal{AS}$ directly issues the \texttt{DEDUP} identifier to the new client to route the data into its cold queue. 
Conversely, if the tag has not been claimed by any other client (i.e., reading a state of $\mathtt{EMPTY}$), $\mathcal{AS}$ grants the data's training right to the new client via a CAS operation. 
This processing mechanism ensures that the joining of new clients does not interrupt the training processes of existing online clients, nor does it require rerunning a global deduplication protocol, effectively achieving the design goal of supporting dynamic client joining.

\end{sloppypar}

\paragraph{Training-right allocation after a disconnection.}
When all online clients complete their training in the current round and proceed to the next, $\mathcal{AS}$ iterates through the bidirectional index and resets the processing state of all data to $\mathtt{EMPTY}$, while retaining the ownership mapping sets and the previous round's trainer records. 
Normally, to ensure system performance and avoid repeated duplicate detection, the training rights for each data item in the new round are retained by default for the previous trainer. 
However, if $U_i$ experienced a disconnection in the previous round, $\mathcal{AS}$ determines that $U_i$ still poses a disconnection risk and adjusts the allocation strategy in the new round. 

When processing duplicate data held by $U_i$, $\mathcal{AS}$ actively intervenes via periodic heartbeat messages. 
Because duplicate data in real-world networks often exhibits complex overlapping ownership relationships (e.g., one data item might be jointly held by $U_i$ and $U_j$, another by $U_i$ and $U_k$, and others simultaneously by three or more clients), the system needs to re-establish a unique trainer for each duplicate data item. 
Specifically, $\mathcal{AS}$ examines the owner set of the tag in the index and selects one online client (e.g., $U_j$) from the remaining clients that do not have a recorded disconnection risk, explicitly excluding $U_i$. 
$\mathcal{AS}$ then encapsulates the \texttt{TRAIN} instruction in $U_j$'s heartbeat response to preferentially allocate the training right. 
Simultaneously, $\mathcal{AS}$ encapsulates the \texttt{DEDUP} identifier in the heartbeat responses of $U_i$ and all other unselected data owners (e.g., $U_k$). 
This operation forces $U_i$ and the unselected clients to skip training this data item and move it into their cold queues. 
Consequently, this history-based allocation mechanism effectively prevents unstable clients from invalidly occupying duplicate data in subsequent rounds, thereby safeguarding the system performance. 

\begin{sloppypar} 
\paragraph{Remark.} 
Our threat model in this paper assumes that clients are honest, and network instability is the sole cause of disconnections. 
The core objective is to verify the concurrency and fault-tolerance performance of the DwT paradigm. 
Therefore, for destructive behaviors where malicious clients claim training rights and repeatedly disconnect, or where they falsely claim that they do not possess certain tags, the introduction of defense mechanisms such as reputation penalties or proofs of data possession can provide effective defense. 
However, defenses against these attacks remain outside the scope of this paper. 
Furthermore, DwT-FL currently utilizes static keys to generate tags. 
To mitigate long-term linkability risks and counter the threat of KS compromise over time, the implementation of epoch-based ephemeral tags or threshold/updatable OPRF schemes (e.g., PIVOT \cite{abs-2608-01390}) serves as an effective approach. 
Nevertheless, these long-term key management mechanisms also fall beyond the scope of this paper, and they remain as directions for future exploration. 

\subsection{Analysis}

\paragraph{Security.}
DwT-FL preserves the confidentiality of each participant's local plaintext data during cross-client duplicate detection, achieving the design goal of data privacy protection. 
When generating protected tags, a client computes a blinded value and runs the OPRF protocol with $\mathcal{KS}$. 
Based on the security of the OPRF protocol, this mechanism ensures that $\mathcal{KS}$ cannot infer the client's input data. 
Meanwhile, $\mathcal{AS}$ only receives the finally computed protected data tags. 
Under the assumption that $\mathcal{AS}$ and $\mathcal{KS}$ do not collude, $\mathcal{AS}$ cannot recover the original data from these tags. 
Furthermore, in line with the standard FL paradigm, raw training data never leaves the local clients, and $\mathcal{AS}$ only receives model parameter updates during the training phase. 
While FL itself may face specific threats (e.g., model inversion \cite{Shi00Z00L25, DengCLLWCLJF26} or data poisoning \cite{TolpeginTGL20}), mitigating these attacks via existing orthogonal defenses (e.g., homomorphic encryption \cite{ChengFJLCPY21, DasuSM22} or secure aggregation \cite{BellBGL020, BonawitzIKMMPRS17}) falls outside the scope of this paper.

\paragraph{Training uniqueness.}
DwT-FL maintains at most one active trainer for each duplicate data item at any time, and ensures that at most one valid training result for that item is committed in each round. 
For any identical plaintext data, the deterministic OPRF mapping generates identical tags. 
Therefore, on $\mathcal{AS}$, all concurrent requests for this data will hit the same entry in the bidirectional index. 
By adopting an optimistic concurrency control strategy combined with atomic CAS operations, the transition process of the state variable from $\mathtt{EMPTY}$ to $\mathtt{PENDING}$ is mutually exclusive. 
This means that under scenarios of highly concurrent preemption by multiple clients, at most one processing thread can successfully modify the state. 
The system subsequently issues the \texttt{TRAIN} identifier to this unique client, thereby eliminating the possibility of the same data being repeatedly trained by multiple clients. 

Furthermore, during failure recovery, $\mathcal{AS}$ unilaterally revokes the disconnected trainer's training right, at which point a new client can take over the data. 
Even if the disconnected client subsequently restores its network connection and attempts to submit delayed computational tasks caused by the disconnection, $\mathcal{AS}$ will directly reject the submission because the data has already been taken over. 
This rejection effectively eliminates the risk of concurrent duplicate training on the same data by both the original and takeover clients. 
\end{sloppypar}

\begin{sloppypar} 
\paragraph{Fault tolerance.}
When a client holding training rights unexpectedly disconnects due to network instability or device failure, its unfinished data temporarily stalls in the $\mathtt{PENDING}$ state. 
Once $\mathcal{AS}$ detects the client's disconnection through the heartbeat timeout mechanism, it directly leverages the inverted user table to rapidly reverse-lookup and locate all unsubmitted tags handled by the disconnected client, safely resetting their states back to $\mathtt{EMPTY}$. 
At this point, other clients holding the same data have already temporarily stored it in their local cold queues according to the initial scheduling. 
Therefore, $\mathcal{AS}$ only needs to issue takeover instructions in these clients' subsequent heartbeat responses. 
Upon receiving these instructions, the clients activate the data in their cold queues and transfer it to the hot queues for supplementary training. 
This mechanism ensures that a local client crash does not result in the loss of training data or a global rollback, providing the system with high fault tolerance. 
\end{sloppypar}

\section{Implementation}
\label{sec:implement}

\begin{sloppypar} 
We implement a prototype of DwT-FL on Linux in Python and C++. 
Entity communication, heartbeat mechanisms, and model training are implemented in Python, whereas the performance-critical bidirectional index and concurrent state-claim mechanism are implemented in C++. 
This design circumvents the synchronization overhead of the Python Global Interpreter Lock and guarantees the atomic semantics of state updates. 
For protected-tag generation, DwT-FL employs a blind OPRF over the Ristretto255 group\footnote{\href{https://ristretto.group/}{https://ristretto.group/}}. 

To eliminate the system overhead introduced by dynamic memory allocation under high-concurrency scenarios, the bidirectional index of $\mathcal{AS}$ (comprising the state table and the inverted user table) is implemented using a C++ pre-allocated memory pool. 
The state table employs an open-addressing hash structure. 
Specifically, the system first computes the FNV-1a hash of the tag to initially locate the state table entry, and performs a full comparison over the 512 hexadecimal characters of the tag during probing to handle hash collisions, ensuring the accuracy of duplicate detection. 

The local training module is built on the PyTorch framework, uses EleutherAI/pythia-14m as a lightweight causal language model for training, and enables TF32 and BF16/FP16 automatic mixed precision and the fused AdamW optimizer. 
To address potential network instability in real-world FL environments, global model parameters are persisted in the Safetensors format and transmitted between clients and $\mathcal{AS}$ using a chunking mechanism with SHA-256 integrity digests. 
This chunking mechanism supports file-offset-based resumable transmission and performs hash verification after data reception completes, thereby preventing incomplete or corrupted model parameter files caused by network failures from being loaded into subsequent aggregation operations. 

When a client uploads its model update, $\mathcal{AS}$ queries the state table to verify the training rights and updates the corresponding data states from $\mathtt{PENDING}$ to $\mathtt{COMMITTED}$. 
Subsequently, $\mathcal{AS}$ executes the aggregation operation only when no data in the $\mathtt{PENDING}$ state exists in the state table. 
Crucially, unlike standard FL that uses a static total count of local records, $\mathcal{AS}$ performs the weighted FedAvg based on the actual number of trained samples. 
Through the state table, $\mathcal{AS}$ directly counts the number of $\mathtt{COMMITTED}$ records for each client to establish aggregation weights. 
For the participant set $\mathcal{S}_t$ in round $t$, let $m_i$ and $w_i^{(t)}$ denote the number of actually trained samples (i.e., the data items in the $\mathtt{COMMITTED}$ state) and the local parameters of the $i$-th client, respectively. 
The global model is computed as:
$w^{(t+1)}=\frac{\sum_{i\in\mathcal{S}_t}m_iw_i^{(t)}}{\sum_{i\in\mathcal{S}_t}m_i}.$ 
\end{sloppypar}
\section{Evaluation}
\label{sec:eval}

In this section, we evaluate DwT-FL by comparing it with the state-of-the-art scheme proposed by Abadi et al. \cite{AbadiDS25} (hereinafter abbreviated as NDSS'25). 

\subsection{Experimental Setup}

We conduct our experiments on a single desktop workstation equipped with an Intel\textsuperscript{\textregistered} Core\texttrademark\ i9-14900K CPU (3.20 GHz, 24 cores, 32 threads), 96 GB of DDR5 memory (95.8 GB available), a 930 GB KIOXIA-EXCERIA PRO SSD, and dual NVIDIA RTX 3090 Ti GPUs. 

\begin{sloppypar} 
\paragraph{Workloads.} 
We use the Haiku dataset\footnote{\href{https://drive.google.com/drive/folders/1SYycnXYaLr4iPeMGxGhtxX1Zs8P_UKLI}{https://drive.google.com/drive/folders/1SYycnXYaLr4iPeMGxGhtxX1Zs8P \allowbreak  \_UKLI}} as the corpus for FL model training. 
The dataset contains 15,281 cleaned and processed short text records, with an average length of approximately 68 characters per record. 
During the preprocessing phase, the system employs deterministic hash partitioning based on the stable record identifier $id$ and a random seed of 17 to allocate 14,532 records to the training set and 749 records to the held-out validation set (5\%). 
Federated training uses the training set.

To evaluate the impact of the number of duplicate records across different clients on DwT-FL, we set a cross-client duplicate ratio $r$. 
For example, given $r=0.3$ and an allocation of $n=1024$ local records per client, the system allocates $1024 \times 0.3 \approx 307$ cross-client duplicate records to each client. 
The system employs a pairwise duplicate allocation strategy to achieve this construction. 
Specifically, records selected as duplicates are allocated to exactly two clients, while non-duplicate records are allocated to a single client. 

\paragraph{Baselines.} 
We compare DwT-FL against the representative baseline system NDSS'25.  
To ensure a fair comparison, rather than evaluating the original open-source implementation of NDSS'25, we re-implement their scheme using the same architecture and technology stack as our prototype. 
Note that existing schemes \cite{AbadiDS25, YeLSZDCYDY25} have demonstrated through evaluations that deduplication can effectively mitigate model memorization risks.
Therefore, we do not additionally evaluate metrics such as model perplexity.
\end{sloppypar}


\subsection{Performance}

\begin{sloppypar} 
For all subsequent experiments, we employ the lightweight causal language model EleutherAI/pythia-14m as the FL model. 
By default, we set the maximum number of backend request processing threads for $\mathcal{AS}$ to 4. 
Furthermore, we introduce a random online delay of 0 to 20 seconds for each client to simulate asynchronous client participation. 
\end{sloppypar}

\begin{sloppypar} 

\begin{figure}[t]
	\centering
	
	\includegraphics[width=0.26\textwidth]{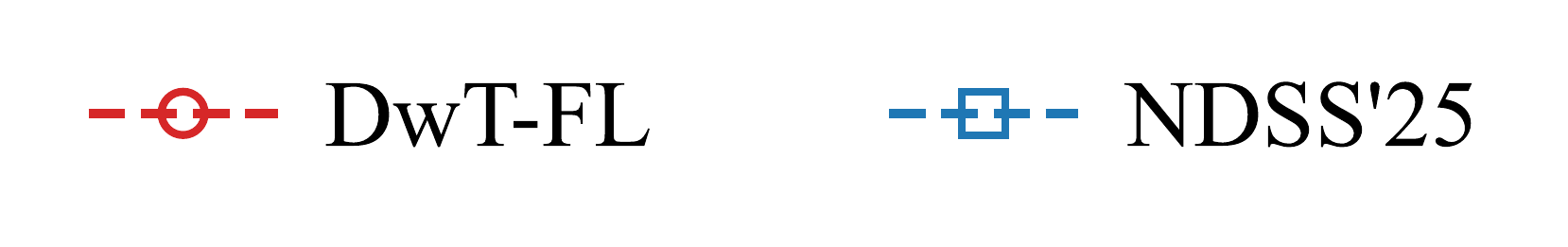} \\
	\vspace{-0.2cm} 
	
	\begin{subfigure}[b]{0.22\textwidth}
		\centering
		\includegraphics[width=\textwidth]{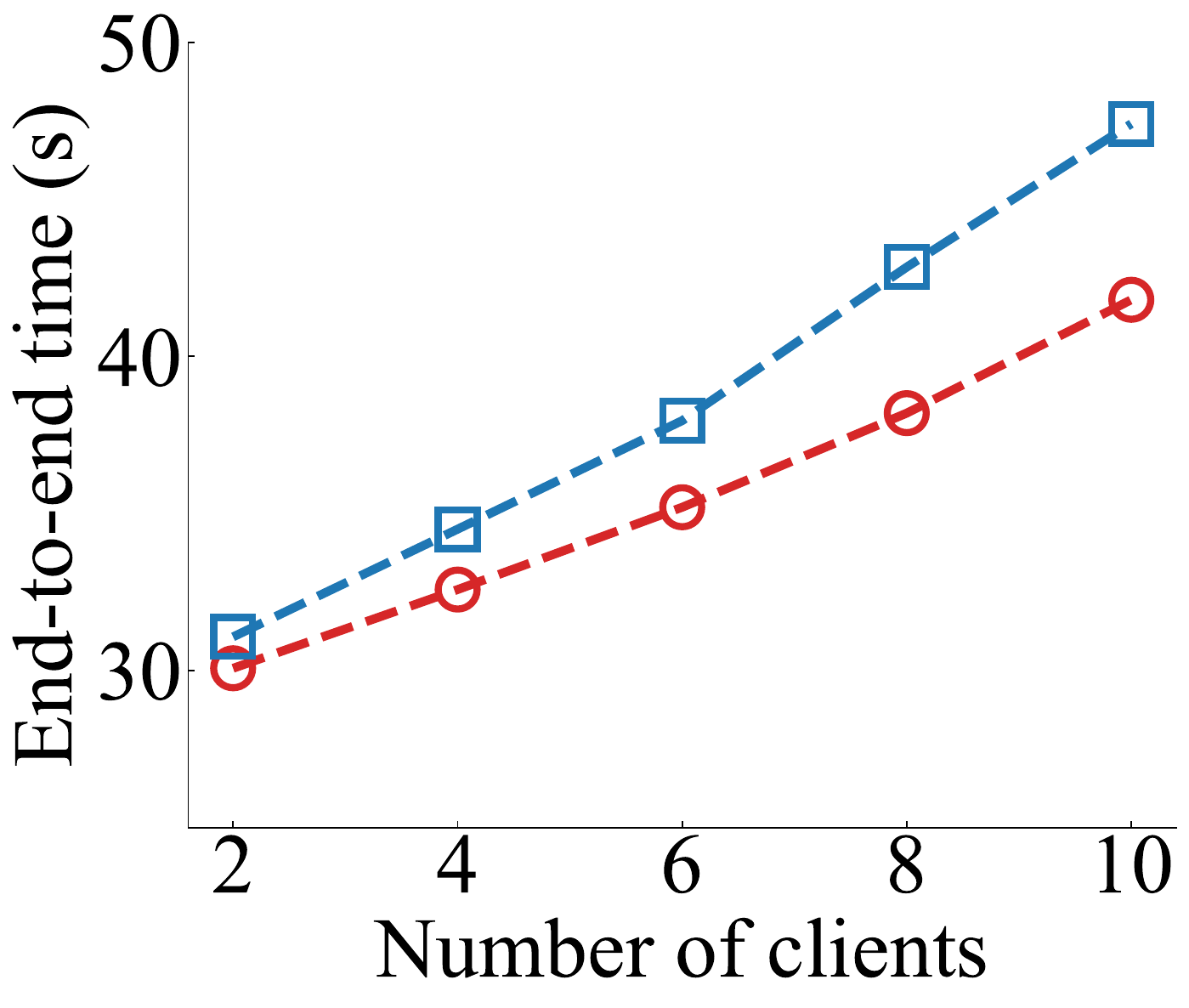}
		\caption{End-to-end time}
		\label{fig:subfig_exp1_1}
	\end{subfigure}
	\hfill 
	\begin{subfigure}[b]{0.22\textwidth}
		\centering
		\includegraphics[width=\textwidth]{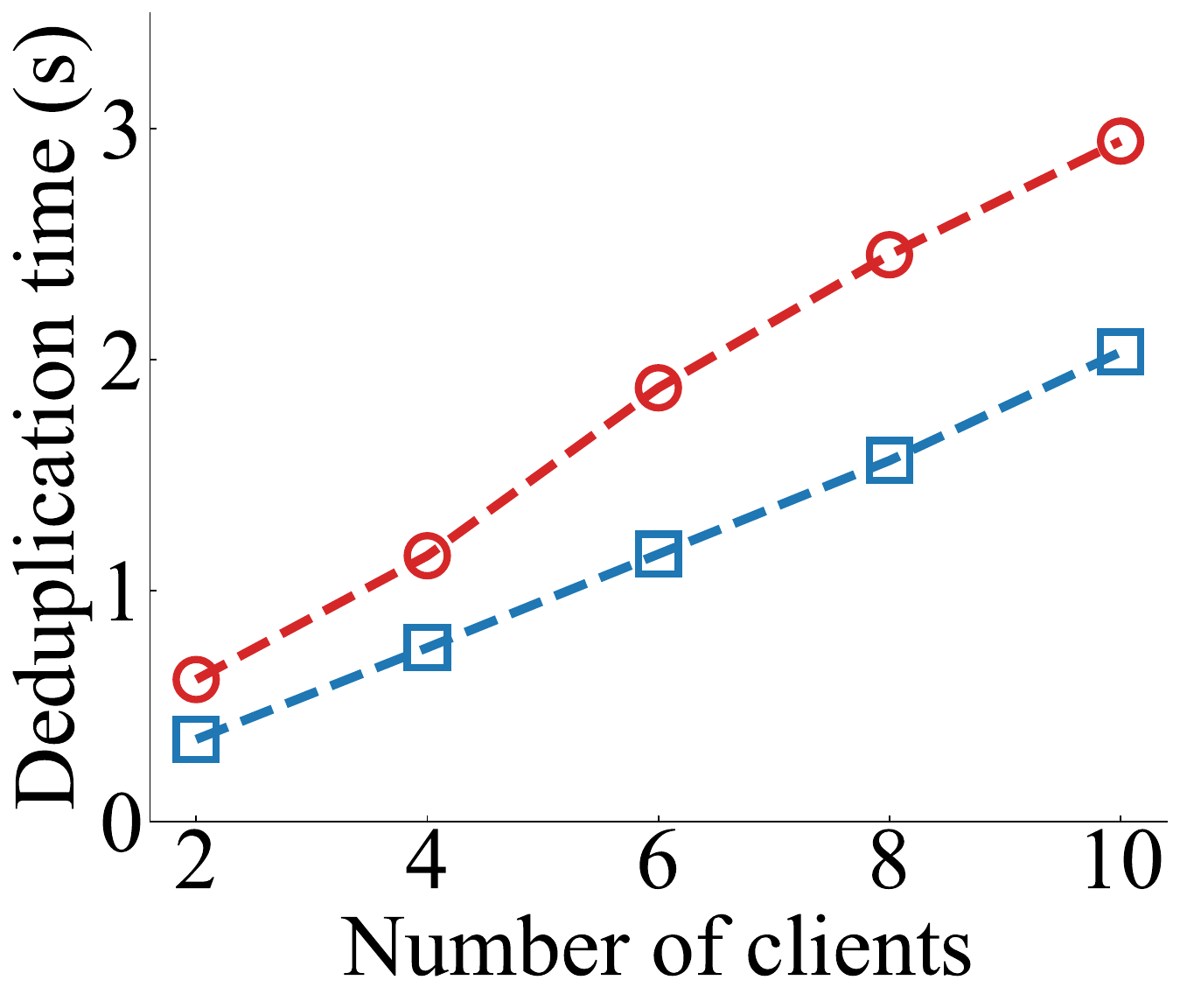}
		\caption{Deduplication time}
		\label{fig:subfig_exp1_2}
	\end{subfigure}
	
	\vspace{0.4cm}
	
	\caption{(Exp\#1) Performance of DwT-FL and NDSS'25 as the number of clients increases.}
	\label{fig:exp1}
	\vspace{-0.5em}
\end{figure}
		
\noindent \textbf{Exp\#1 (Impact of the number of clients).} 
To evaluate the scalability of DwT-FL, we set the number of clients to 2, 4, 6, 8, and 10. 
We allocate 1,024 local records to each client and set the cross-client duplicate ratio to $r=0.3$. 
Under this configuration, each client contains approximately 307 duplicate records shared pairwise with another client. 

We define the end-to-end time starts when the first client initiates a data deduplication request, and ends upon the completion of global model distribution.
As shown in Fig. \ref{fig:subfig_exp1_1}, as the number of clients increases from 2 to 10, the total number of records held by the clients increases from 2,048 to 10,240, and the end-to-end completion times of both DwT-FL and NDSS'25 exhibit an approximately linear growth trend.  
Note that the total number of records here refers to the total number of records allocated to the clients, which includes cross-client duplicate records, and is not equivalent to the number of globally unique records. 
The end-to-end time of DwT-FL is lower than that of NDSS'25, and this advantage becomes more pronounced as the number of clients increases. 
This is because DwT-FL executes deduplication and training in parallel, and this parallel execution reduces the impact of asynchronous client participation. 

Furthermore, the tested deduplication phase encompasses the interactive workflow from tag generation and $\mathcal{AS}$ index updates to CAS-based concurrent training right preemption and local queue construction. 
The deduplication time of NDSS'25 is measured from the moment all clients come online until all clients complete duplicate detection. 
It should be noted that to ensure a fair comparison of deduplication time, we exclude the client wait time in NDSS'25. 
As shown in Fig. \ref{fig:subfig_exp1_2}, the deduplication time of DwT-FL is slightly higher than that of NDSS'25. 
This is because DwT-FL needs to additionally execute index updates, concurrent training-right claiming, and the construction of local client queues, and these additional operations lead to a higher duplicate detection time. 
However, because the deduplication time accounts for an extremely low proportion of the end-to-end time, this overhead is acceptable. 
For example, under the 10-client configuration, the end-to-end time is 41.80 s, whereas the deduplication time is 2.95 s. 

\begin{figure}[t]
	\centering
	\begin{minipage}{0.48\linewidth}
		\centering
		\includegraphics[width=\linewidth]{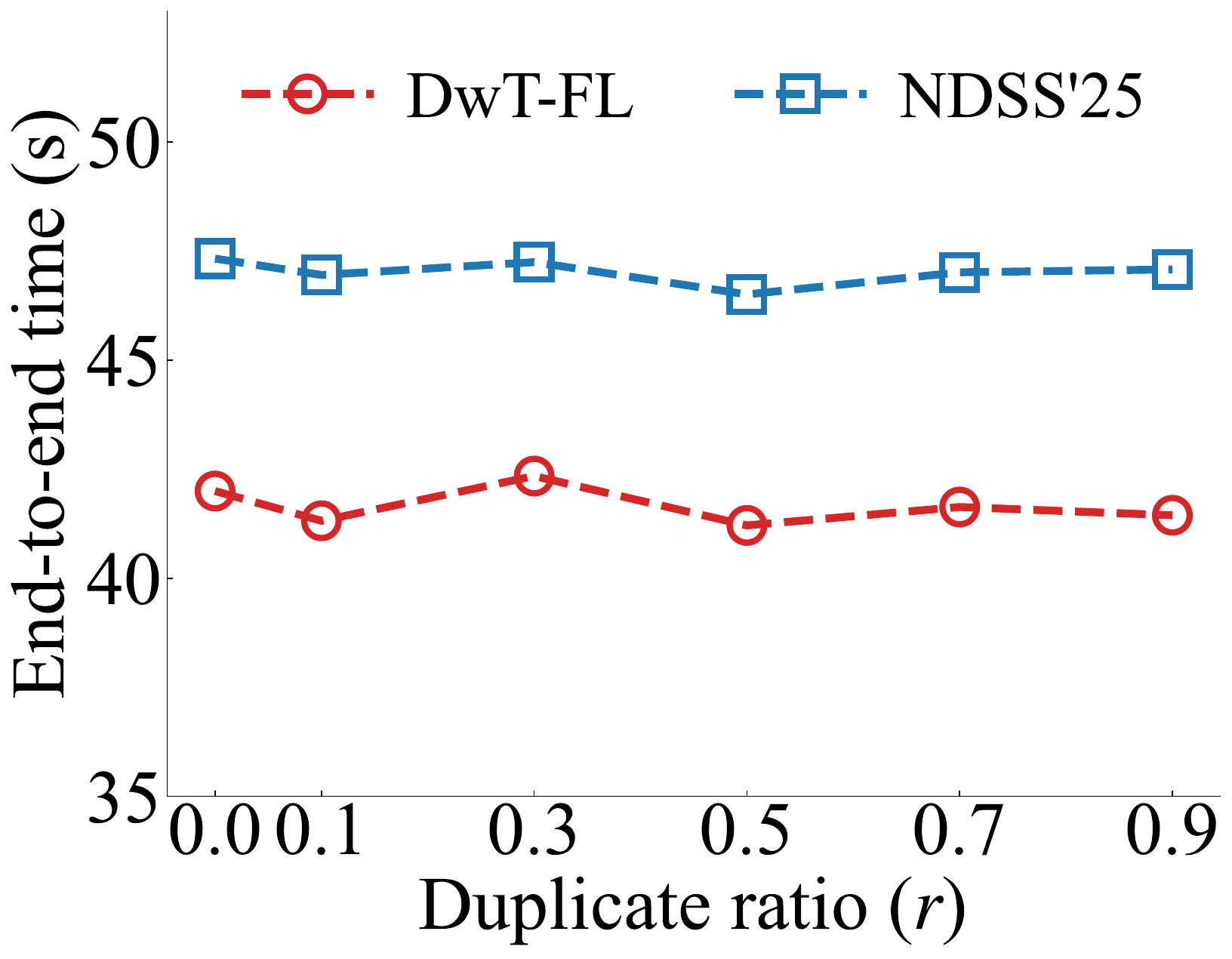}
		\caption{(Exp\#2) Impact of duplicate ratio}
		\label{fig:exp2}
	\end{minipage}
	\hfill
	\begin{minipage}{0.48\linewidth}
		\centering
		\includegraphics[width=\linewidth]{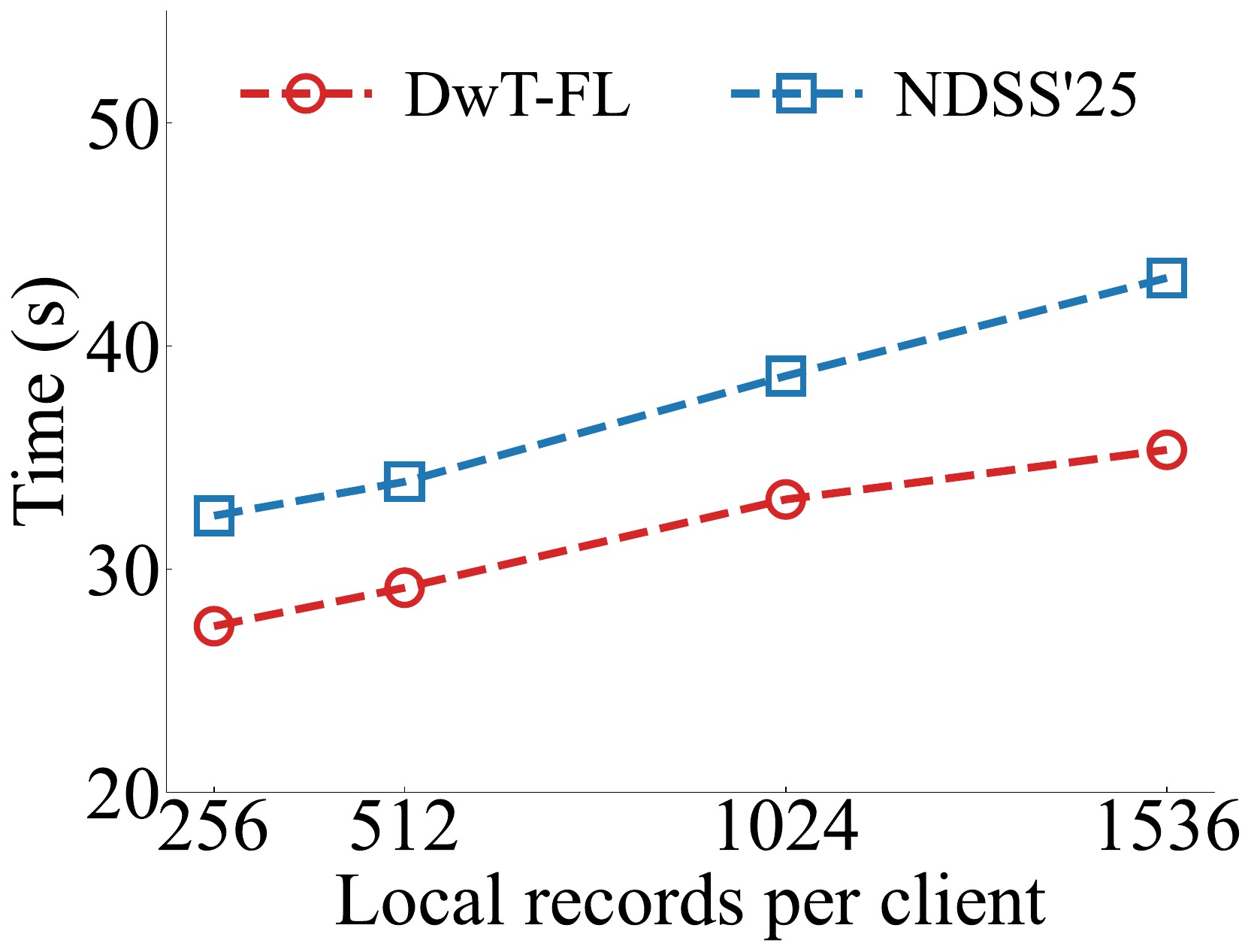}
		\caption{(Exp\#3) Impact of data scale}
		\label{fig:exp3}
	\end{minipage}
	\vspace{-0.1cm}
\end{figure}

\noindent \textbf{Exp\#2 (Impact of duplicate ratio).}
To evaluate the impact of the cross-client duplicate ratio on DwT-FL, we set the duplicate ratio $r$ to 0.0, 0.1, 0.3, 0.5, 0.7, and 0.9. 
Under a system scale of 10 clients, we allocate a fixed 1,024 local records to each client, resulting in a total of 10,240 records across all clients. 
As shown in Fig. \ref{fig:exp2}, when the duplicate ratio $r$ increases from 0.0 to 0.9, the end-to-end time of DwT-FL fluctuates between 41.21 s and 42.34 s. 
Because DwT-FL executes deduplication and training in parallel, the end-to-end time is lower than NDSS'25. 
The end-to-end time of the system is primarily limited by the total number of input records. 
Because this total number of input records remains fixed at 10,240 under different values of $r$ in our setup, the time metrics do not change significantly as the duplicate ratio increases. 
It should be noted that the metadata overhead of DwT-FL does not change with the duplicate ratio, but only increases as the data number grows. 

\end{sloppypar}

\begin{sloppypar} 
	
\noindent \textbf{Exp\#3 (Impact of data scale).}
To evaluate the impact of data scale on the performance of DwT-FL, we fix the system to include 10 clients, set the cross-client duplicate ratio to $r=0.3$, and set the number of local records allocated to each client to 256, 512, 1,024, and 1,536. 
Correspondingly, the total number of input records across all clients is 2,560, 5,120, 10,240, and 15,360, respectively. 
As shown in Fig. \ref{fig:exp3}, as the data scale per client increases from 256 to 1,536 records, the end-to-end time of DwT-FL grows from 27.44 s to 35.35 s, and the end-to-end time of NDSS'25 grows from 32.39 s to 43.05 s. 
The end-to-end time of DwT-FL remains consistently lower than that of NDSS'25. 
This is because DwT-FL can execute local training in parallel during the wait time for other clients to come online. 
This approach overlaps the deduplication time and the training time, and consequently it reduces the overall end-to-end time. 

\begin{figure}[t]
	\centering
	
	\includegraphics[width=0.26\textwidth]{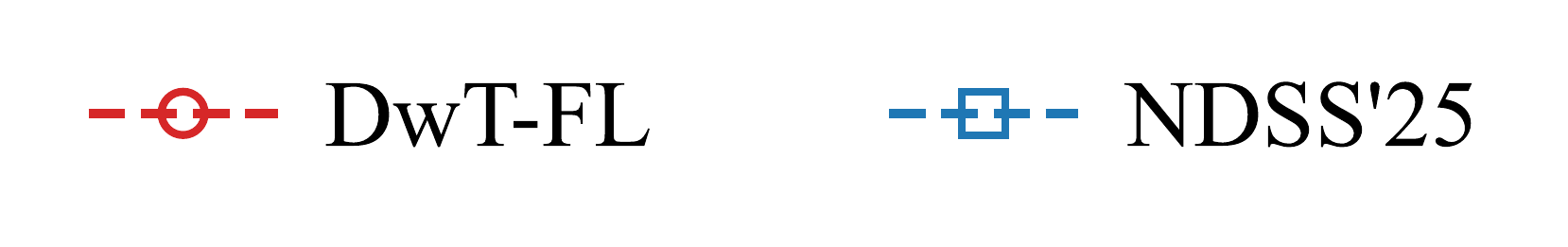} \\
	\vspace{-0.2cm} 
	
	\begin{subfigure}[b]{0.22\textwidth}
		\centering
		\includegraphics[width=\textwidth]{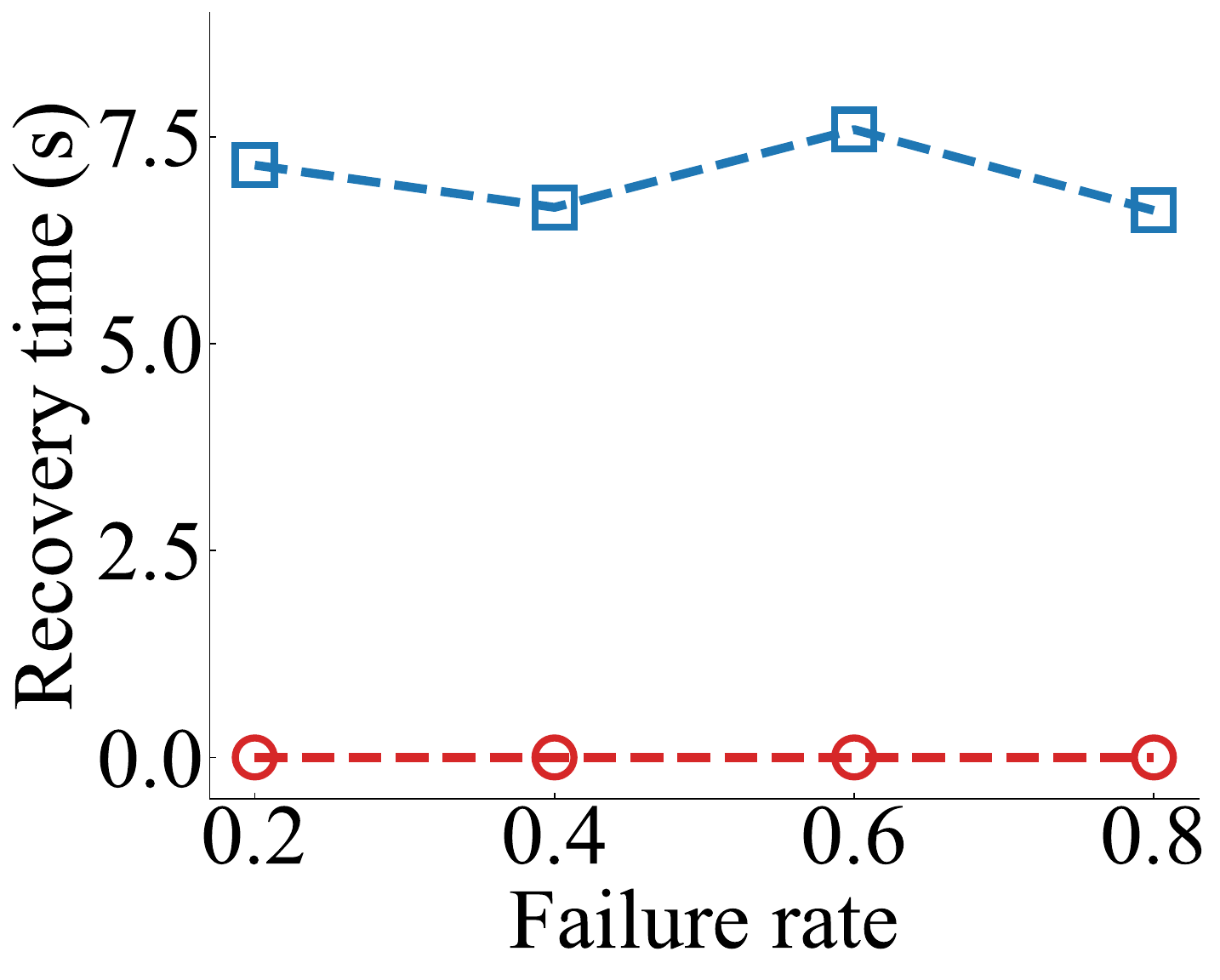}
		\caption{Deduplication phase}
		\label{fig:subfig_exp4_1}
	\end{subfigure}
	\hfill 
	\begin{subfigure}[b]{0.22\textwidth}
		\centering
		\includegraphics[width=\textwidth]{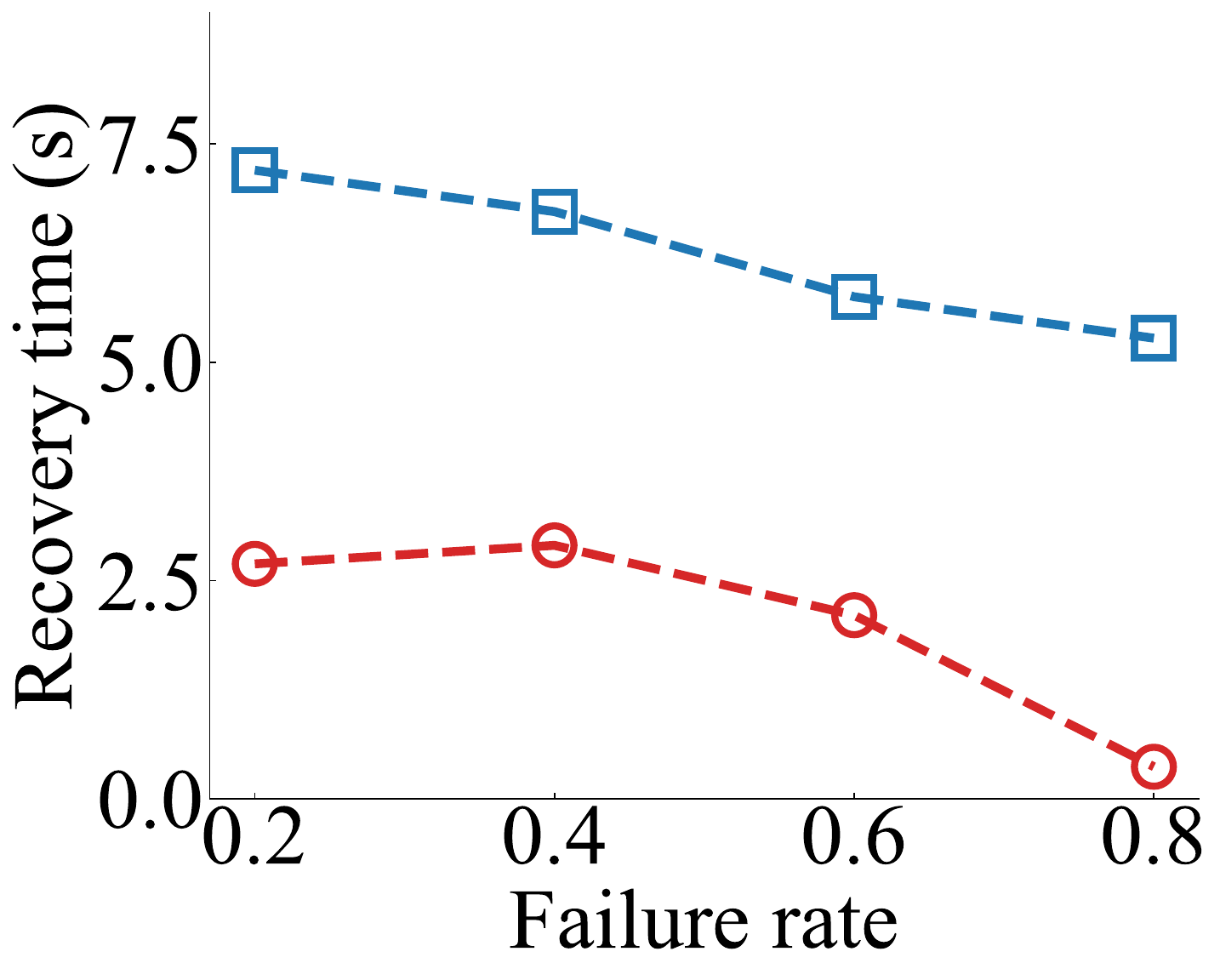}
		\caption{Training phase}
		\label{fig:subfig_exp4_2}
	\end{subfigure}
	
	\vspace{0.4cm}
	
	\caption{(Exp\#4) Failure recovery.}
	\label{fig:exp4}
	\vspace{-0.5em}
\end{figure}

\noindent \textbf{Exp\#4 (Performance of failure recovery).}
To evaluate the efficiency of the failure recovery in DwT-FL, we fix the system configuration at 10 clients, set the cross-client duplicate ratio to $r=0.3$, and set the failure rates to 0.2, 0.4, 0.6, and 0.8. 
The failure rate represents the ratio of the number of disconnected clients to the total number of clients. 
We randomly force clients to disconnect to simulate unexpected client disconnects in real-world FL. 
We evaluate the failure recovery time of DwT-FL and NDSS'25. 
The failure recovery time starts when the first client disconnects, and ends when other clients complete the takeover of the duplicate data from the disconnected clients. 
It should be noted that we do not use the end-to-end time as the evaluation metric. 
This is because the end-to-end time includes the time for local model training, model update uploads, and global model aggregation. 
The time consumed by these processes affects the actual performance differences in failure recovery between different schemes, and therefore it cannot accurately reflect the efficiency of the failure recovery mechanism in DwT-FL. 

As shown in Fig. \ref{fig:subfig_exp4_1}, the failure recovery time of DwT-FL in the deduplication phase is consistently 0. 
This is because client disconnects during the deduplication phase do not affect the execution of DwT-FL. 
In contrast, NDSS'25 incurs a higher failure recovery time. 
As shown in Fig. \ref{fig:subfig_exp4_2}, during the training phase, the failure recovery time of DwT-FL is significantly lower than that of NDSS'25, reducing the failure recovery time by up to $93.04\%$. 
This is because DwT-FL can execute failure recovery in parallel during training, and does not need to re-execute the duplicate detection protocol. 
Furthermore, when the failure rate is 0.8, the failure recovery time decreases for both DwT-FL and NDSS'25. 
This occurs because only 2 clients remain online, which significantly reduces the amount of recoverable data. 
Consequently, the number of data requiring training right reallocation drops substantially, leading to a shorter failure recovery time.

\begin{figure}[t]
	\centering
	
	\includegraphics[width=0.26\textwidth]{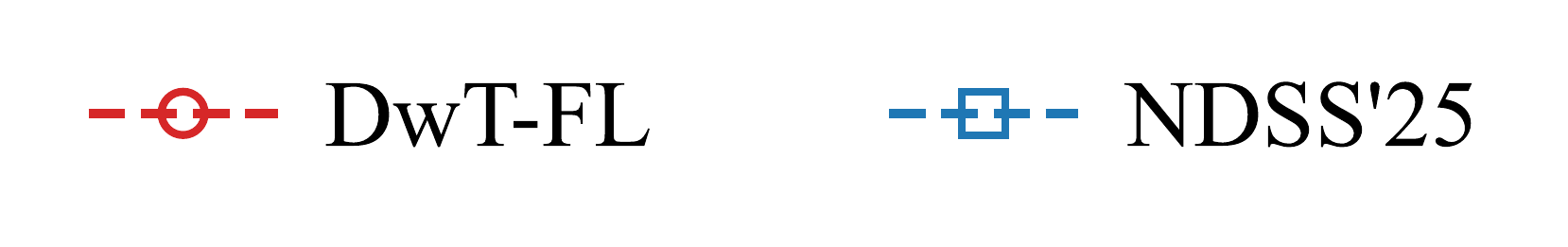} \\
	\vspace{-0.2cm} 
	
	\begin{subfigure}[b]{0.22\textwidth}
		\centering
		\includegraphics[width=\textwidth]{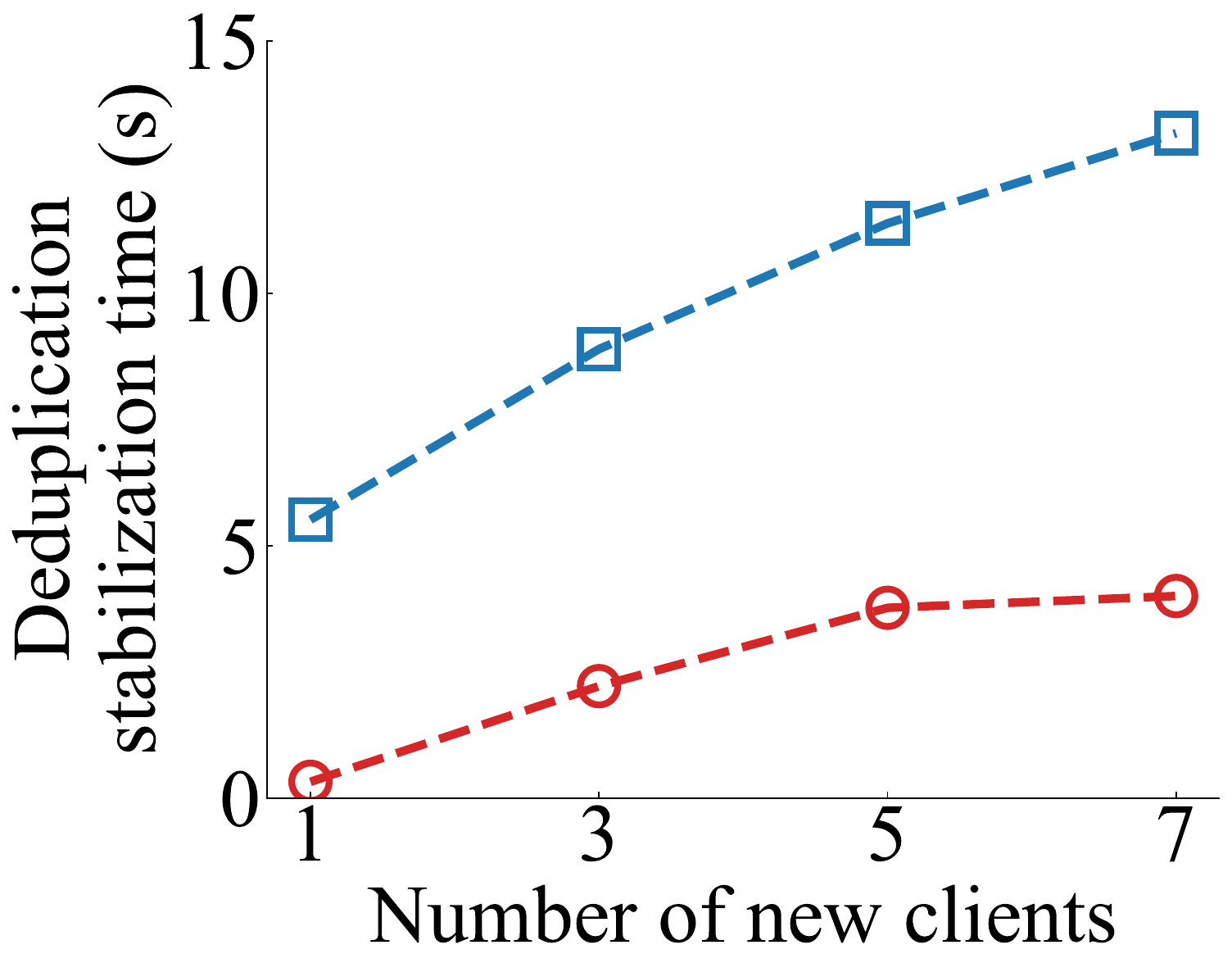}
		\caption{Deduplication phase}
		\label{fig:subfig_exp5_1}
	\end{subfigure}
	\hfill 
	\begin{subfigure}[b]{0.22\textwidth}
		\centering
		\includegraphics[width=\textwidth]{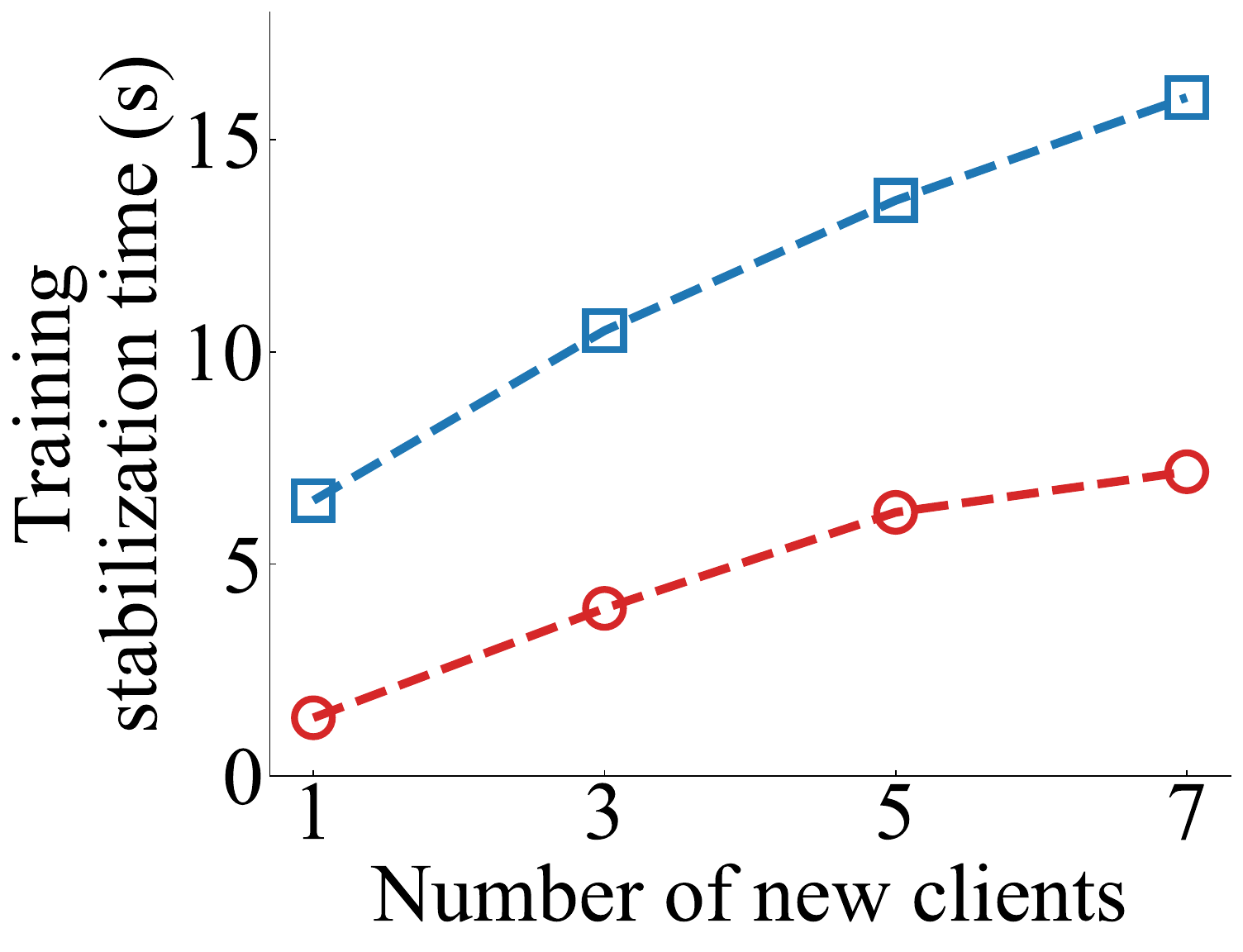}
		\caption{Training phase}
		\label{fig:subfig_exp5_2}
	\end{subfigure}
	
	\vspace{0.4cm}
	
	\caption{(Exp\#5) Dynamic client joining.}
	\label{fig:exp5}
	\vspace{-0.5em}
\end{figure}

\noindent \textbf{Exp\#5 (Performance of dynamic client joining).}
To evaluate the efficiency of DwT-FL under dynamic client joining, we fix the initial system configuration at 10 clients, set the cross-client duplicate ratio to $r=0.3$, and set the number of newly joining clients to 1, 3, 5, and 7. 
We allow clients to join randomly to simulate client joining scenarios in real-world FL. 
Specifically, we generate a random wait time for each new client, and the client joins the FL process after waiting for this duration. 
We separately evaluate client joining during the deduplication phase and the training phase. 
For client joining in the deduplication phase, we evaluate the deduplication stabilization time. 
The deduplication stabilization time starts when the first client joins, and ends when all clients complete duplicate detection. 
For client joining in the training phase, we evaluate the training stabilization time. 
The training stabilization time starts when the first client joins, and ends when all clients begin training. 
For NDSS'25, we adopt a straightforward method to support client joining. 
Specifically, when a new client joins, it separately executes duplicate detection with the existing clients, and then it proceeds to the training phase. 

As shown in Fig. \ref{fig:subfig_exp5_1}, under the scenario of client joining in the deduplication phase, the deduplication stabilization time of DwT-FL is significantly lower than that of NDSS'25, and it reduces the deduplication stabilization time by up to $94.18\%$. 
This is because newly joined clients in DwT-FL only need to perform incremental comparisons with the tags stored on the $\mathcal{AS}$ to quickly complete deduplication, whereas NDSS'25 needs to re-execute the global deduplication protocol or perform complex global state synchronization. 
Similarly, as shown in Fig. \ref{fig:subfig_exp5_2}, the training stabilization time of DwT-FL is significantly better than that of NDSS'25, and it achieves a maximum time reduction of $78.87\%$. 
This is because new clients in DwT-FL can directly execute duplicate detection with the tags stored in the $\mathcal{AS}$, and they do not affect the training of existing clients. 
Furthermore, it can be observed that as the number of newly joined clients increases, the time consumption for both schemes increases. 
However, the time growth curve of DwT-FL is flatter, and its time reduction ratio compared to NDSS'25 consistently remains above 54\%. 
This demonstrates the efficiency of DwT-FL under dynamic client joining.

\begin{figure}[t]
	\centering         
	\includegraphics[width=0.6\linewidth]{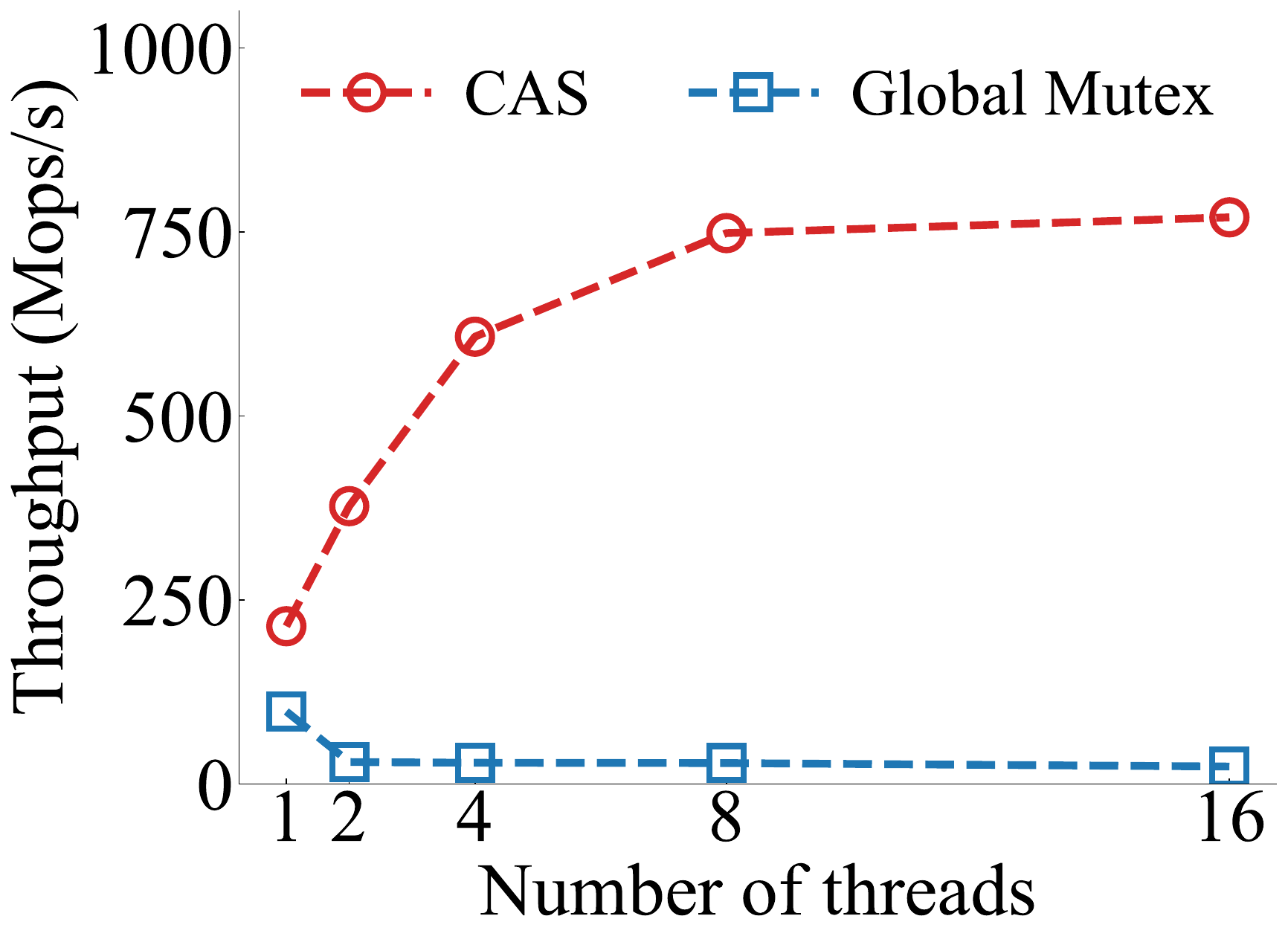} 
	\caption{(Exp\#6) Ablation study.}
	\label{fig:Exp6}   
\end{figure}

\noindent \textbf{Exp\#6 (Ablation study on DwT-FL).}
To evaluate the effectiveness of the proposed components, namely CAS-based training-right claiming and failure recovery based on bidirectional index, we design ablation studies. 
Specifically, we replace CAS with a global Mutex in the training-right claiming, replace the bidirectional index with a full table scan in the failure recovery, and disable the history-based allocation mechanism. 

For CAS-based training-right claiming, we configure 10 concurrent threads to compete simultaneously for the training rights of 1,024 data records, and ultimately only one thread successfully obtains the training right for each record.
When the system adopts CAS, the average time to complete this process is 0.121 ms, and the throughput is 85.27 Mops/s (million operations per second). 
When the system adopts a global Mutex, the average time is 0.408 ms, and the throughput is 25.09 Mops/s.
Consequently, CAS reduces the processing time by 70.58\% and increases the throughput by 3.40 times. 
Furthermore, to evaluate the scalability under different thread counts, we set the number of concurrent threads to 1, 2, 4, 8, and 16. 
In this setup, each thread executes 50,000 state transitions from $\mathtt{EMPTY}$ to $\mathtt{PENDING}$ on the data. 
As shown in Fig. \ref{fig:Exp6},the throughput of CAS rises from 214.36 Mops/s to 769.91 Mops/s, whereas the throughput of the global Mutex decreases from 98.62 Mops/s to 23.75 Mops/s. 
Under the 16-thread configuration, the throughput of CAS reaches 32.41 times that of the Mutex. 
These results demonstrate that CAS effectively improves the system performance. 

To evaluate the failure recovery based on bidirectional index, we conduct the experiment under a configuration of 10 clients, 1,024 records per client, and a duplicate ratio of $r=0.3$. 
We force one client to disconnect and release its 307 duplicate data. 
When the system adopts the bidirectional index, the time to locate and release the relevant data training rights is 1.992 ms, whereas a full table scan requires 12.964 ms. 
We also evaluate the failure recovery time. 
This metric spans from the moment the system detects the disconnected client to the moment other clients complete the takeover of all duplicate data from that client. 
Within this process, $\mathcal{AS}$ needs to locate the tags corresponding to the data trained by the disconnected client and set their states to $\mathtt{EMPTY}$. 
The failure recovery times for the bidirectional index and the full table scan are 75.092 ms and 85.951 ms, respectively. 

\begin{figure}[t]
	\centering
	
	\includegraphics[width=0.5\textwidth]{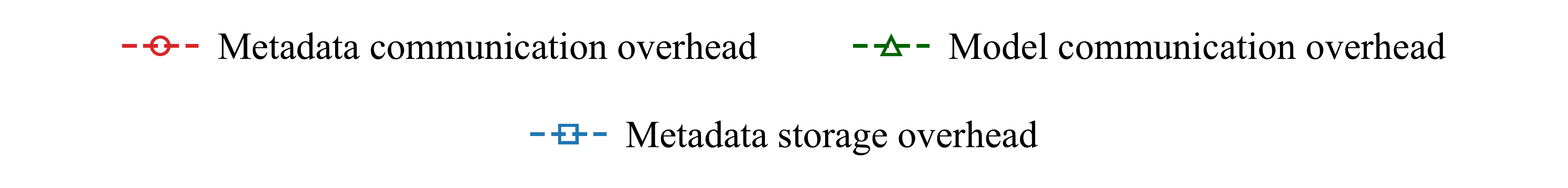} \\
	
	\begin{subfigure}[b]{0.23\textwidth}
		\centering
		\includegraphics[width=\textwidth]{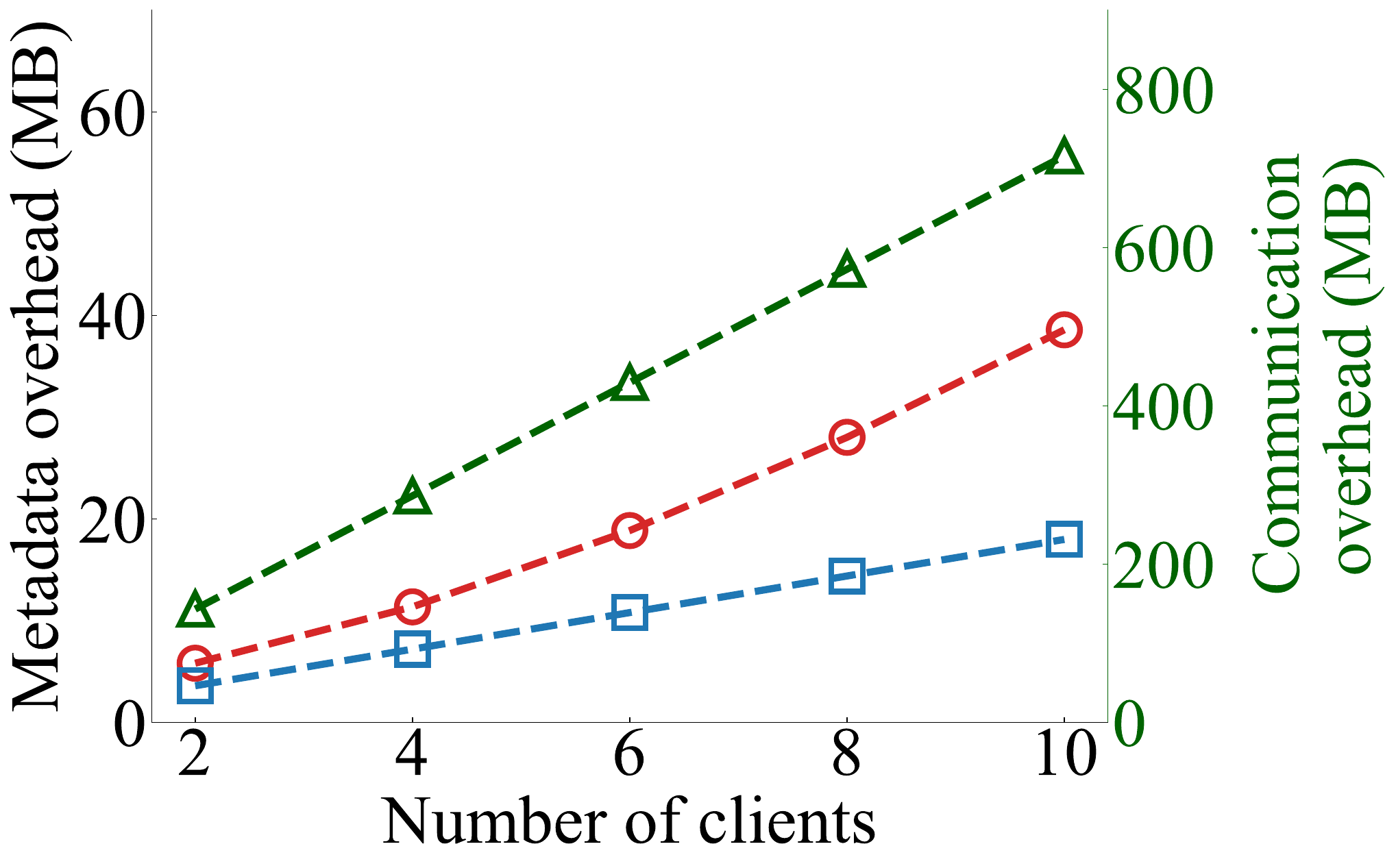}
		\caption{Number of clients}
		\label{fig:subfig_exp7_1}
	\end{subfigure}
	\hfill 
	\begin{subfigure}[b]{0.23\textwidth}
		\centering
		\includegraphics[width=\textwidth]{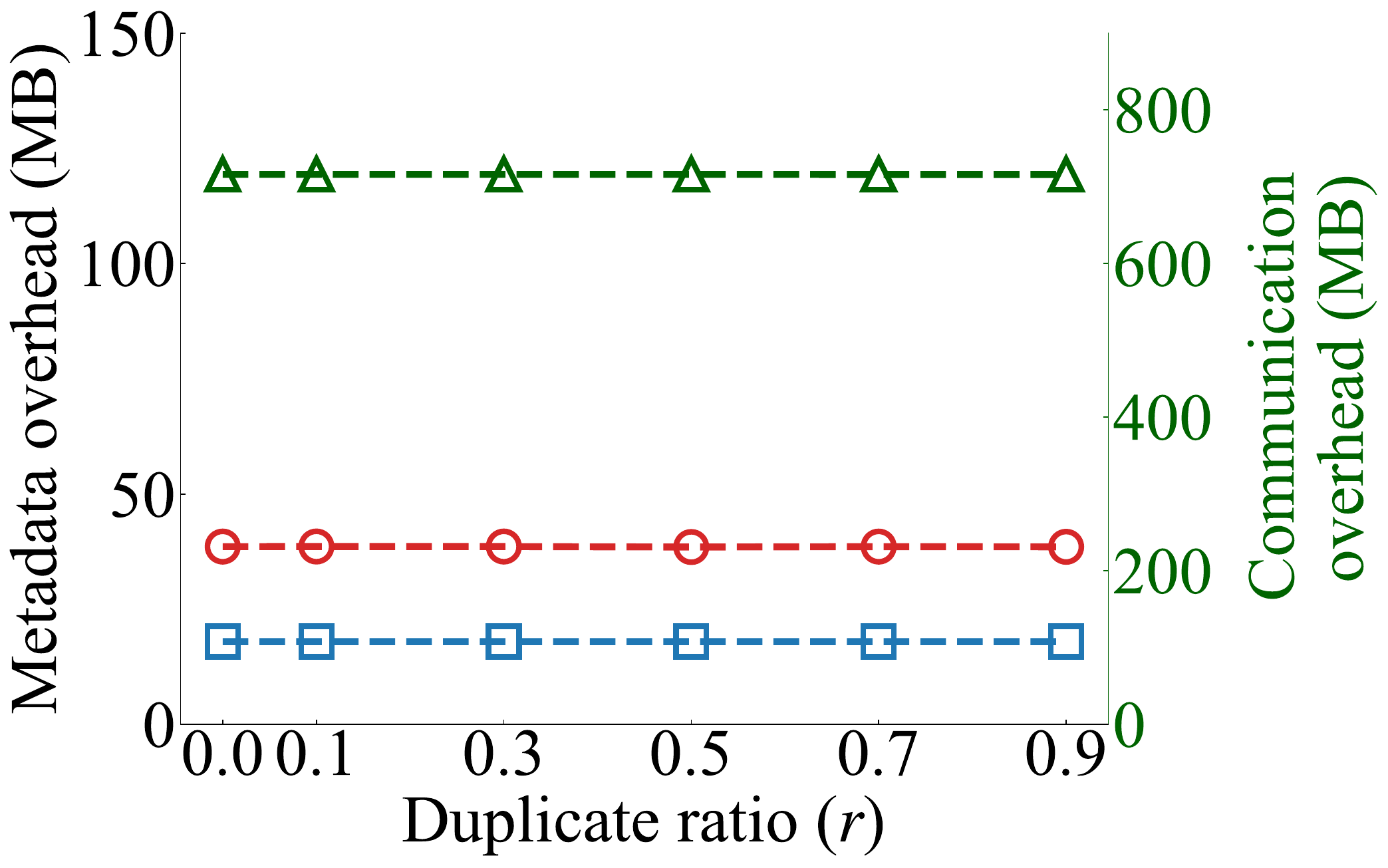}
		\caption{Duplicate ratio}
		\label{fig:subfig_exp7_2}
	\end{subfigure}
	
	\vspace{0.4cm}
	
	\caption{(Exp\#7) Metadata overhead.}
	\label{fig:exp7}
	\vspace{-0.5em}
\end{figure}

\noindent \textbf{Exp\#7 (Metadata overhead).}
Regarding communication and storage overhead, cross-client deduplication introduces certain metadata overhead. 
This overhead includes two categories: 
1) \textit{communication metadata} generated during network transmission (e.g., duplicate detection tags and heartbeat control packets); 
2) \textit{storage metadata} required for state maintenance (e.g., the bidirectional index on $\mathcal{AS}$ and the locally persisted mapping tables on clients). 

We separately evaluate how the metadata overhead varies with the number of clients (which also reflects its variation with the data scale) and the duplicate ratio. 
As shown in Fig. \ref{fig:subfig_exp7_1}, when the number of clients increases from 2 to 10, the metadata communication overhead of DwT-FL rises from 5.818 MB to 38.593 MB. 
Correspondingly, the model parameter transmission volume in a single FL round scales from 143.192 MB to 715.971 MB. 
Consequently, the ratio of the metadata communication overhead to the model transmission load remains extremely low, and it only experiences a marginal increase from 4.06\% to 5.39\%.
This proportion will further decrease as the model size increases.
This demonstrates that the duplicate detection step of DwT-FL will not become a communication bottleneck for FL. 
In terms of storage, when the system scales to 10 clients, the metadata storage overhead reaches 17.989 MB, and the raw model parameter payload for a single round is 562.878 MB. 
The metadata storage overhead accounts for only 3.20\% of this raw model payload. 
Furthermore, the metadata storage overhead exhibits a linear growth trend with the increase in the number of clients.
When the system expands from 2 to 10 clients, the metadata storage overhead increases from 3.604 MB to 17.989 MB. 
These results prove that DwT-FL imposes no significant communication or storage overhead on $\mathcal{AS}$ and the clients, thereby exhibiting favorable scalability.  

As shown in Fig. \ref{fig:subfig_exp7_2}, the metadata overhead does not change with the duplicate ratio. 
This is because regardless of whether the data is duplicated, clients need to execute duplicate detection for every local record. 
Consequently, the metadata communication overhead depends on the total input number (10,240 records) rather than the deduplicated data number. 
Meanwhile, the metadata storage overhead is related to the local mappings on clients and the size of the $\mathcal{AS}$ index. 
Since clients do not delete duplicate data, the number of local mappings on clients is not affected by the duplicate ratio. 
Additionally, the size of the $\mathcal{AS}$ index is affected by the pre-allocated storage capacity of the index (a strategy we designed in our implementation), and the data scale affects this pre-allocated storage capacity. 
The duplicate ratio does not affect the data scale. 
Therefore, the size of the $\mathcal{AS}$ index remains unchanged.  

\end{sloppypar}

\section{Conclusion}
\label{sec:conclusion}

This paper proposes a novel DwT paradigm to address the high fault-tolerance costs and the lack of support for dynamic client joining in prior privacy-preserving deduplication schemes in FL. 
DwT transforms cross-client deduplication in FL from a one-time, globally synchronous preprocessing operation into a continuous online service with state management, concurrent claiming, and failure recovery. 
To support this paradigm, we design the DwT-FL scheme. 
DwT-FL enables the parallel execution of privacy-preserving cross-client data deduplication and client-local training through a CAS-based concurrent state claim mechanism and a hot-cold dual-queue scheduling strategy. 
Experimental evaluations demonstrate that DwT-FL significantly reduces the overhead of failure recovery and dynamic client joining.

\bibliographystyle{ACM-Reference-Format}
\bibliography{bib}


\end{document}